\documentclass{article}

\usepackage[bookmarks=true]{hyperref}
\usepackage{graphicx,epsfig,color}
\usepackage{cite}
\usepackage{amsmath}
\numberwithin{equation}{section}
\usepackage[english]{babel}
\usepackage{amssymb,amsfonts,amsmath}
\usepackage{verbatim}
\usepackage{tikz}
\usepackage{bbold}
\newcommand{\cA}{{\cal A}}  \newcommand{\cB}{{\cal B}}
  \newcommand{\cD}{{\cal D}}
  \newcommand{\cF}{{\cal F}}

  \newcommand{\cL}{{\cal L}}
  \newcommand{\cN}{{\cal N}}

  \newcommand{\cV}{{\cal V}}
  
\newcommand{\cY}{{\cal Y}}

\newcommand{\be}{\begin{equation}} \newcommand{\ee}{\end{equation}}
\newcommand{\bea}{\begin{eqnarray}} \newcommand{\eea}{\end{eqnarray}}
\newcommand{\beann}{\begin{eqnarray*}}  \newcommand{\eeann}{\end{eqnarray*}}
\newcommand{\bfig}{\begin{figure}} \newcommand{\efig}{\end{figure}}
\newcommand{\ba}{\begin{array}} \newcommand{\ea}{\end{array}}
\newcommand{\bcen}{\begin{center}} \newcommand{\ecen}{\end{center}}
\newcommand{\btab}{\begin{tabular}} \newcommand{\etab}{\end{tabular}}

\newcommand{\vev}[1]{\left\langle{#1}\right\rangle}

\newtheorem{Proposition}{Proposition}[section]

\newtheorem{Theorem}{Theorem}[section]
\newtheorem{Lemma}{Lemma}[section]
\newtheorem{Corrolary}{Corrolary}[section]

\newcommand{\bp}{\begin{Proposition}}   \newcommand{\ep}{\end{Proposition}}
\newcommand{\bt}{\begin{Theorem}}   \newcommand{\et}{\end{Theorem}}
\newcommand{\bl}{\begin{Lemma}}     \newcommand{\el}{\end{Lemma}}
\newcommand{\bc}{\begin{Corrolary}} \newcommand{\ec}{\end{Corrolary}}

\def\ep{\epsilon}

\title{\bf Baryon Superfluids in AdS/CFT with Flavor}
\author{Carlos Hoyos${}^{1,a}$, Georgios Itsios${}^{1,2,b}$, Orestis Vasilakis${}^{1,c}$\\
$\ $\\
${}^1$ {\em Department of Physics, Universidad de Oviedo}\\
{\em Avda.~Calvo Sotelo 18, ES-33007 Oviedo, Spain}\\
$\ $\\
${}^2$ {\em Instituto de F\'{\i}sica Te\'orica, UNESP-Universidade Estadual Paulista,}\\
{\em R. Dr. Bento T. Ferraz 271, Bl. II, Sao Paulo 01140-070, SP, Brazil}\\
$\ $\\
{\small ${}^a$\tt{hoyoscarlos@uniovi.es}, ${}^b$\tt{gitsios@gmail.com}, ${}^c$\tt{vasilakisorestis@uniovi.es}}\\ 
}
\date{}

\begin{document}


{\twocolumn[
\begin{flushright}
FPAUO-16/14 \\
\end{flushright}
\maketitle
  \begin{@twocolumnfalse}
  \maketitle
    \begin{abstract}
\noindent Baryonic matter is notoriously difficult to deal with in the large-N limit, as baryons become operators of very large dimension with N fields in the fundamental representation. This issue is also present in gauge/gravity duals as baryons are described by very heavy localized objects. There are however alternative large-N extrapolations of QCD where small baryonic operators exist and can be treated on an equal footing to mesons. We explore the possibility of turning on a finite density of ``light'' baryons in a theory with a hadronic mass gap using a gauge/gravity construction based on the D3/D7 intersection. We find a novel phase with spontaneous breaking of baryon symmetry at zero temperature. 
      \vspace{0.5cm}
    \end{abstract}
  \end{@twocolumnfalse}
]}
\onecolumn

\tableofcontents

\section{Introduction}

It is notoriously difficult to describe from first principles dense baryonic matter in QCD at small temperatures and large densities. Perturbation theory can be used only at extremely high densities \cite{Kurkela:2009gj,Kurkela:2016was}. Lattice calculations on the other hand are restricted to values of the baryon chemical potential smaller than the temperature \cite{deForcrand:2010ys}. One then has to rely on phenomenological models, but those are usually fitted to describe the physics in very different regimes, so it is far from clear that they can give an accurate description (see e.g. section 7.3 of \cite{Brambilla:2014jmp} for a review). A common difficulty is that models are usually adapted to describe either quark or hadronic matter, but there should be a transition (which could be smooth) between the two as the density is changed.

It would be very interesting to have a theory that can be described in all density regimes from first principles. A natural proposal is to study gauge/gravity models \cite{Maldacena:1997re,Gubser:1998bc,Witten:1998qj}. A strongly coupled field theory in the large-$N_c$ limit has a holographic dual description in terms of a classical higher dimensional geometry, which is a black hole if the temperature is nonzero \cite{Witten:1998zw}. Flavor degrees of freedom are introduced by adding D-branes in the geometry (we will refer to them as ``flavor branes'') \cite{Karch:2002sh}. If the number of flavors is much smaller than the number of colors $N_f\ll N_c$, the branes can be treated as probes. Flavor currents are dual to gauge fields living in the worldvolume of the branes, in such a way that a state with finite density is realized by having non-zero electric flux on the brane. 

The worldvolume electric flux has to be sourced by some charges, which we can think of as open strings attached to the brane. In the models that are usually considered, the two string endpoints carry opposite charges, so in order to have a non-zero density one of the endpoints should end on the brane and the other somewhere else. A possibility is that the strings extend from the flavor brane to the horizon. One can think of this situation as having quarks in a plasma. A finite density of them will pull the brane embedding, in such a way that the finite density state can be described by a brane embedding that reaches the horizon from where the electric field is sourced \cite{Kobayashi:2006sb}. 

The second possibility is that the strings extend from the flavor branes to a different kind of brane, dual to a ``baryon vertex'', which wraps the internal directions in the geometry and is point-like in the field theory directions \cite{Witten:1998xy,Brandhuber:1998xy}. The tension of the strings produces a force such that typically at the equilibrium configuration the brane dual to the baryon vertex lies on the flavor brane and can be described as a solitonic configuration on the flavor brane worldvolume \cite{Sakai:2004cn,Hata:2007mb,Seki:2008mu}. This corresponds to baryonic matter in the dual field theory. 

Although holographic models can accommodate both quark and baryonic matter in this fashion, there is a clear asymmetry between the two. In order to describe baryonic matter one needs to study multi-soliton solutions \cite{Rho:2009ym,Kaplunovsky:2012gb,Bolognesi:2013jba,Kaplunovsky:2015zsa} or use a phenomenological approach if one is interested in homogeneous states \cite{Bergman:2007wp,Rozali:2007rx,Kim:2007zm,deBoer:2012ij,Ghoroku:2012am,Li:2015uea,Preis:2016fsp,Elliot-Ripley:2016uwb,Elliot-Ripley:2016ctk}, with the drawback that the physical properties of the state depends on the assumptions one needs to make. Furthermore, stable soliton solutions have sizes that are typically of the order of the string scale \cite{Hata:2007mb}, thus casting doubts on the validity of the brane action used to find those solutions. Therefore, our understanding of baryonic matter in holographic models is on much more shaky ground than that of quark matter.

The difference between the quark and baryonic matter in holographic models can be traced back to the large-$N_c$ limit. Baryonic operators are constructed with $N_c$ fundamental fields, thus they are very heavy objects and this is reflected in the holographic dual description, where they are described by branes or solitonic configurations in the flavor branes. Mesonic operators on the other hand can be constructed with a small number of fields and have a holographic dual description in terms of open strings ending on the flavor branes, or small fluctuations of the fields living in their worldvolume. This hierarchy between baryons and mesons is an artifact of the large-$N_c$ limit and is not observed in real QCD (see e.g. \cite{Sonnenschein:2016pim}). It is thus desirable to study different models where this distinction is erased.

The holographic picture gives us a clue about how to do this. Mesons are open strings attached to the flavor branes. They have zero baryon number because the endpoints of each string have opposite charge. This suggests that one could describe states with nonzero baryon number with open strings if both endpoints had the same charge, but this would be possible only if there are unoriented strings. Therefore, in order to describe baryons on equal footing as mesons (we will refer to them as ``light baryons'') it is necessary to introduce orientifold planes in the geometry. In this paper we will consider a particular model proposed in \cite{HoyosBadajoz:2009hb}, which is an orientifold of the well-known D3/D7 model. We will introduce a chemical potential in the phase with hadronic bound states and show that there is a critical value above which the baryon density becomes non-zero. This provides the first example of homogeneous baryonic matter in a holographic model that does not rely on additional assumptions on the gravity side besides the usual weak coupling and small curvature approximations.

In the new phase light baryon operators acquire an expectation value. Thus in this phase baryon symmetry is spontaneously broken and the holographic model describes a baryon superfluid state. To our knowledge this is the first realization of a baryon superfluid phase in a string theory model. Baryon superfluid matter may exist in nuclei and the interior of neutron stars, as originally shown by Migdal \cite{Migdal:1960}. At asymptotically large densities, the color flavor locking (CFL) phase exhibits baryon superfluidity \cite{Alford:1998mk}.\footnote{CFL phases in a model with flavor branes were introduced in \cite{Chen:2009kx}, however those still preserve a $U(1)$ baryon symmetry.} In both cases weak attractive interactions between charged fermions (nucleons or quarks) induce the formation of a superfluid following the standard Bardeen-Cooper-Schrieffer (BCS) theory. However, in different density regimes interactions become stronger and a there can be a crossover to a state where fermions bound in molecules and form a Bose-Einstein condensate (BEC) (see e.g. \cite{He:2013gga} for a review). The baryon superfluid described by the holographic model resembles such BEC states.

The outline of the paper is as follows. In \S~\ref{sec:model} we review the field theory model with light baryon operators and its holographic dual. In \S~\ref{sec:action} we derive the effective action for the fields dual to light baryonic and mesonic operators from the flavor brane action. In \S~\ref{sec:density} we study the phase diagram at zero temperature and non-zero chemical potential, and show that the zero charge density phase becomes thermodynamically disfavored above a critical value of the chemical potential. In \S~\ref{sec:ground} we construct a superfluid baryon phase in a simple case with a single scalar operator condensing. We present our conclusions and discuss future directions in \S~\ref{sec:conclusions}. We have collected several technical details in the Appendices.

\section{A holographic model with light baryons}\label{sec:model}

The model was originally introduced in \cite{HoyosBadajoz:2009hb}, based on the usual $D3/D7$ intersection \cite{Karch:2002sh}. Let us review first the usual case. The theory is based on the following arrangement of $D3$ and $D7$ branes, where bullets denote the directions along which the branes are extended
$$
\begin{array}{|c|cccccccccc|}
\hline                        & 0        & 1         & 2        & 3         & 4          & 5        & 6        & 7        & 8        & 9  \\
\hline D3                  & \bullet & \bullet & \bullet & \bullet &             &           &           &            &           &    \\
\hline D7            & \bullet & \bullet & \bullet & \bullet &             &           & \bullet & \bullet & \bullet & \bullet \\  \hline
\end{array}
$$
On the field theory side there are $2 N_c$ $D3$ branes (the ``color branes'') whose low-energy description is $U(2N_c)$ $\cN=4$ super Yang-Mills (SYM).\footnote{We introduce factors of $2$ for later convenience when we introduce orbifold and orientifold projections.} In $\cN=2$ language it contains a vector multiplet and a hypermultiplet in the adjoint representation. In $\cN=1$ language there is a vector multiplet $V$ and three chiral multiplets in the adjoint representation. We can split the space transverse to the $D3$ branes into three planes $45$, $67$ and $89$, in such a way that the expectation value of the scalar component of each chiral multiplet is associated to the position of $D3$ branes on each plane. We thus label the corresponding chiral multiplets as $X_{45}$, $X_{67}$ and $X_{89}$. In this notation $X_{45}$ is charged under rotations on the $45$ plane while the other chiral multiplets are neutral.

In the model there are also $2 N_f$  $D7$ branes  (the ``flavor branes'') that introduce the same number of hypermultiplets in the fundamental representation. Flavor hypermultiplets contain two chiral multiplets $Q$ and $\tilde{Q}$ in conjugate representations. From the point of view of the theory living in the $(3+1)$-dimensional intersection of $D3$ and $D7$ branes there is a global $U(2N_f)$ symmetry, a $U(1)_R$ symmetry acting as rotations on the $45$ plane and a $SO(4)\simeq SU(2)_L\times SU(2)_R$ acting as rotations on the $6789$ space. The $R$-symmetry group  is $SU(2)_R\times U(1)_R$. 

The holographic dual to the theory on the $D3$ branes when $N_c\to \infty$ is type IIB string theory in an $AdS_5\times S^5$ geometry with RR 5-form flux $\cF_5=F_5+{}^*F_5$, $F_5=dC_4$
\begin{equation}
ds^2=\frac{r^2}{R^2} dx_{1,3}^2+\frac{R^2}{r^2} dr^2+R^2d\Omega_5^2, \ \ C_4=\frac{R^4}{r^4} dx^0\wedge dx^1\wedge dx^2\wedge dx^3.
\end{equation} 
The $AdS$ radius $R$ is related to the 't Hooft coupling of the $\cN=4$ SYM theory as $R^4/(\alpha')^2=\lambda_{YM}$, where $\ell_s=\sqrt{\alpha'}$ is the string length. 

If the number of flavors is much smaller than the number of colors $N_f\ll N_c$, the $D7$ branes do not affect the physics of the $D3$ branes to leading order in the large-$N_c$ expansion and can be treated as probes. The flavored degrees of freedom on the field theory side are captured by the dynamics of probe $D7$ branes embedded in the $AdS_5\times S^5$ geometry. These are extended along the $AdS_5$ directions and wrap a $S^3\subset S^5$. We will use the following conventions for indices for the fields on the branes
\begin{itemize}

  \item $a$, $b$, ... : denote the worldvolume directions of the D7 brane, 
  
  \item $\mu$, $\nu$, ... : denote the spacetime directions along which lies the D7 brane,
  
  \item $i$, $j$, ... : denote the spacetime directions that are transverse to the D7 brane,
  
  \item $A$, $B$, ... : denote the directions of the D7 brane that wrap $S^3$,
  
  \item and $M$, $N$, ... : denote the directions of the D7 brane along the spacetime directions 0, 1, 2, 3 and $\rho$.

\end{itemize}
So for the worldvolume directions of the D7 brane we have $a = \left(M,A\right)$.

The bosonic fields on the $D7$ brane are a $U(2N_f)$ gauge field, $A_a$, and scalars in the adjoint representation of flavor that are associated to the directions transverse to the $D7$ brane, $X^i$, $i=4,5$. Supersymmetric embeddings have a very simple form in a particular set of coordinates.  Following \cite{Kruczenski:2003be}, we will write the metric as
\begin{equation}
ds^2=\frac{r^2}{R^2} dx_{1,3}^2+\frac{R^2}{r^2} ((dY^4)^2+(dY^5)^2+d\rho^2+\rho^2d\Omega_3^2),
\end{equation} 
where $r^2=(Y^4)^2+(Y^5)^2+L^2$. Identifying $x^M=\{x^0,x^1,x^2,x^3,\rho\}$ and the coordinates of the $S^3$ with the worldvolume coordinates of the $D7$, the solutions to the equations for the embedding (up to rigid rotations in the $(Y^4, Y^5)$ plane) are simply $X^4=Y^4=L$, $X^5=Y^5=0$. The induced metric on the flavor brane can then be written as
\begin{equation}\label{eq:flavmet}
ds^2=\frac{\rho^2+L^2}{R^2} dx_{1,3}^2+\frac{R^2}{\rho^2+L^2} \left(d\rho^2+\rho^2d\Omega_3^2\right).
\end{equation}

To obtain light baryon operators this model is deformed by the introduction of an orbifold acting as a reflection in the directions along the $D7$ branes that are transverse to the $D3$ branes and an orientifold $O7$ plane parallel to the D7 branes. The new brane configuration is:
$$
\begin{array}{|c|cccccccccc|}
\hline                        & 0        & 1         & 2        & 3         & 4          & 5        & 6        & 7        & 8        & 9  \\
\hline D3                  & \bullet & \bullet & \bullet & \bullet &             &           &           &            &           &    \\
\hline O7/D7            & \bullet & \bullet & \bullet & \bullet &             &           & \bullet & \bullet & \bullet & \bullet \\  
\hline \mathbb{Z}_2 & \bullet & \bullet & \bullet & \bullet &  \bullet & \bullet &           &            &            &  \\ \hline
\end{array}
$$
For the orbifold we have $\mathbb{Z}_2\subset SU(2)_L$ and thus $\cN=2$ supersymmetry (SUSY) is preserved. The effect of the orbifold and the orientifold can be seen as imposing a condition over the physical states of the theory. One can define an orbifold/orientifold action in the original theory that consists of a reflection, on the $6789$ directions for the orbifold and the $45$ directions for the orientifold, plus transformations acting on the Chan-Paton factors of the open strings ending on the branes.  These transformations can be different for the $D3$ and $D7$ branes and for the orientifold they also involve a change in the orientation of the strings. In both cases the orbifold/orientifold action squares to the identity. Physical states are invariant under both the orbifold and orientifold action and this projects out a set of states of the original theory. The orbifold action we are considering has no fixed point, so there is no additional twisted sector. Since the orbifold and orientifold actions act on Chan-Paton factors, they can change the gauge group and the representation of the fields on the branes. 

The combined effect of the orbifold and the orientifold is to reduce by half the rank of the color and flavor groups, to $U(N_c)$ and $U(N_f)$ respectively, and to turn the $\cN=2$ adjoint hypermultiplet (constructed with the chiral multiplets $X_{67}$, $X_{89}$) into a hypermultiplet in a two-index antisymmetric representation of $U(N_c)$ (details can be found in \cite{HoyosBadajoz:2009hb}). In $\cN=1$ notation, the antisymmetric chiral multiplets $A_{[\alpha\beta]}$, $\widetilde{A}^{[\alpha\beta]}$ can be combined with the (anti)fundamental chiral multiplets $Q_\alpha$, $\widetilde{Q}^\alpha$ to construct BPS gauge invariant operators with non-zero baryonic charge
\begin{equation}
\cB=Q\widetilde{A}Q, \ \ \widetilde{\cB}=\widetilde{Q}A\widetilde{Q}.
\end{equation}
Since $Q$, $A$ are bosonic operators, the light baryon operators above have to be in the antisymmetric representation of the $SU(N_f)$ flavor group. Note that when $N_c=3$ the antisymmetric representation coincides with the anti-fundamental,\footnote{Since they are related by the invariant antisymmetric symbol $A_{[\alpha\beta]}=\epsilon_{\alpha\beta\gamma}\tilde{q}^\gamma$.} so the color structure of $\cB$ and $\widetilde{\cB}$ is exactly the same as that of baryons in QCD.

On the gravity side the geometry is changed to $AdS_5\times  \mathbb{RP}^5$ and the $N_f$ probe $D7$ branes wrap a $\mathbb{RP}^3\subset \mathbb{RP}^5$. The theory on the $D7$  branes can be obtained by projecting out the modes in the original setup that are not invariant under the combined orbifold and orientifold action. The projection acts differently depending on the angular momentum $\ell$ of the fields along the $S^3$ directions, since the orbifold action acts as a reflection in these directions. We can distinguish between the gauge field components along the $x^M$ directions, $A_M$, the gauge field components along the $S^3$ directions, $A_A$ and the scalar fields associated to the directions transverse to the $D7$ branes $X^i$, $i=4,5$ (we will omit the index for the rest of this section and refer to both as $X$) . The orbifold and orientifold projections are determined by the following matrices acting on the Chan-Paton factors of the $D7$ open strings 
\begin{equation}
\gamma_7= i\sigma_3 \otimes \mathbb{1}_{N_f},\ \ 
\omega_7=\sigma_1\otimes \mathbb{1}_{N_f}\, .
\end{equation}
Physical states must satisfy the following conditions ($T$ denotes transpose)
\begin{equation}\label{eq:projections}
\begin{array}{ll}
A_M^\ell = (-1)^\ell \gamma_7 A_M^\ell\gamma_7^{-1}, &  A_M=-\omega_7 (A_M^\ell)^T\omega_7^{-1},\\[5pt]
A_A^\ell = (-1)^{\ell+1}\gamma_7 A_A^\ell\gamma_7^{-1}, &  A_A=- \omega_7 (A_A^\ell)^T\omega_7^{-1},\\[5pt]
X^\ell = (-1)^\ell \gamma_7 X^\ell\gamma_7^{-1}, &  X=- \omega_7 (X^\ell)^T\omega_7^{-1}.
\end{array}
\end{equation}
Note that the action of the orbifold projection depends on the angular momentum on the $S^3$. This can be implemented by a large gauge transformation on the $S^3$, such that the transformed fields are periodic in the $\mathbb{RP}^3$ orbit of the $\mathbb{Z}_2$ action that identifies the antipodal points on the $S^3$.  We give the details in Appendix~\ref{app:orbifold}. 
The combined orbifold plus orientifold projections produce the following structure
\begin{equation}\label{eq:projected}
\begin{split}
&A_M^{\rm even} \sim X^{\rm even}\sim A_A^{\rm odd} \sim \left(\begin{array}{cc}  H & \\ & -H^* \end{array} \right),\\[5pt]
&A_M^{\rm odd} \sim X^{\rm odd}\sim A_A^{\rm even} \sim \left(\begin{array}{cc}   & B \\ -B^* & \end{array} \right),\\
\end{split}
\end{equation}
Where $H=H^\dagger$ is Hermitian and $B=-B^T$ is antisymmetric, both $N_f\times N_f$ matrices. The first type are dual to operators in the adjoint representation of $U(N_f)$ flavor, while the second are dual to the baryon operators in the antisymmetric representation. Note that they are charged under the $U(1)_B$ symmetry, since
\begin{equation}\label{eq:projectedfields}
i\left[ \left(\begin{array}{cc}  \mathbb{1} & \\ & -\mathbb{1} \end{array} \right) \,,\, \left(\begin{array}{cc}   & B \\ -B^* & \end{array} \right) \right]=2  \left(\begin{array}{cc}   & iB \\ -(iB)^* & \end{array} \right) .
\end{equation}
The spectrum  of BPS operators follows from the spectrum of the original theory specified in \cite{Kruczenski:2003be}.  The multiplets of BPS operators are classified by the representation of the $SU(2)_L$ group, which can be labelled by a half-integer $j_L$. The representation under $SU(2)_R\times U(1)_R$ is determined by a half-integer $j_R$ and an integer $R$. We will group all these using the notation $(j_R,j_L)_R$. In addition, there is a $U(N_f)$ global flavor symmetry under which hadronic operators can be in the adjoint (adj) or two-index antisymmetric (antis) representations. The lowest BPS operators, corresponding to relevant and marginal operators, are dual to the following modes ($X_{45}$ is an adjoint chiral multiplet)
$$
\begin{array}{cccccc}
\Delta & \ell & \text{mode}  & \text{$U(N_f)$ rep} & (j_R, j_L)_R & \text{operator (candidates)}\\
2 & 1 & A_A & \text{adj} & \left( 1,0\right)_0 & Q\tilde{Q}\\
3 & 0 & X     & \text{adj} & \left( 0,0\right)_2 & QX_{45} \tilde{Q}\\
3 & 0 & A_M & \text{adj} & \left( 0,0 \right)_0  & J_\mu \sim Q D_\mu \tilde{Q} \\
3 & 2 & A_A & \text{antis} & \left( \frac{3}{2},\frac{1}{2}\right)_0  & \cB\sim Q\tilde{A}Q\\
4 & 1 & X    & \text{antis} & \left( \frac{1}{2},\frac{1}{2}\right)_2  & Q\tilde{A}X_{45}Q \\
4 & 1 & A_M & \text{antis} & \left( \frac{1}{2},\frac{1}{2}\right)_0  & Q \tilde{A} D_\mu Q \\
4 & 3 & A_A &  \text{adj} & \left( 2,1\right)_0 & QD^2\tilde{Q} 
\end{array}
$$

\section{Effective action of fields dual to light baryons}\label{sec:action}

In order to do the projections we introduce two stacks of $N_f$ D7 branes, one at $Y^4=L$ and the other at $Y^4=-L$. 
In terms of the scalar fields on the brane  $X^4=L\, \sigma^3\otimes \mathbb{1}_{N_f}$, $X^5=0$, which is of the form \eqref{eq:projected}. The induced metric and the coupling to the RR potential in each stack is the same, so there is no difference between the flavor embeddings before and after the projection at zero density. When the density is non-zero we expect that the shape of the embedding will be affected. In order to avoid complications we will use the approximation that the density is small enough to neglect this effect, so we will expand around a background configuration with the induced metric (\ref{eq:flavmet}).

The lowest BPS operators with flavor charges have a dual description in terms of the fields on the brane, whose dynamics are determined by the non-Abelian D-brane action. Although the general form of the action is not known, in the supersymmetric case Myers used T-duality arguments to propose an action \cite{Myers:1999ps}. The bosonic fields living on the D-brane worldvolume are the gauge field $A_a$, whose field strength is $F_{ab}$ and the scalar fields $X^i$, all in the adjoint representation of $U(2N_f)$. Let us define $\lambda= 2\pi \alpha'$, $\sigma^a$ the coordinates in the D-brane worldvolume and the worldvolume scalar fields will be expanded as $X^i(\sigma)=x^i(\sigma)+\lambda \Phi^i$ with $x^i$ the position of the branes in the transverse directions. The tension of the Dp-brane is
\begin{equation}
T_7=\frac{2\pi}{g_s(2\pi \lambda)^4}. 
\end{equation}
The D-brane action has a Born-Infeld and a Wess-Zumino term\footnote{We choose the same normalizations as \cite{Myers:1999ps} so the BI and WZ terms have the same overall factor.} 
\begin{equation}
\begin{split}
S_{D7}=& S_{BI}+S_{WZ},
\\[5pt]
S_{BI}=&-T_7 \int d^{8}\sigma \, {\rm Tr}_{2N_f}\Bigg[e^{-\phi}\sqrt{-\det\Big(P\big[ E_{ab}+E_{ai}(Q^{-1}- \delta)^{ij}E_{jb}\big]+\lambda F_{ab}\Big)\det Q^i_{\ j} }\Bigg],
\\[5pt]
S_{WZ}= & T_7 \int_{D7}  \, {\rm Tr}_{2N_f}\Bigg[ P\Big(e^{\frac{i }{\lambda} i_ X i_X}   \sum_n C_n\wedge  e^B \Big)\wedge e^{\lambda F}\Bigg].
\end{split}
\end{equation}
In these expressions enter the closed string fields
\begin{equation}
E_{\mu\nu}=G_{\mu\nu}+B_{\mu\nu}, \ \  {Q^i}_ j={\delta^i}_j+\frac{i}{\lambda}\left[X^i,X^k \right]E_{kj}\, ,
\end{equation}
where $G_{\mu\nu}$ is the string frame metric, $B_{\mu\nu}$ is the NS two-form field and $P[\cdots]$ denotes the pullback
\begin{equation}
P[E_{ab}]=E_{\mu\nu}\partial_a x^\mu \partial_b x^\nu + E_{\mu i} \partial_{a} x^\mu D_{b} X^i+E_{i\mu} D_{a} X^i \partial_{b} x^\mu + E_{ij}D_a X^i D_b X^j, \ \ D_a X^i=\partial_a X^i+i[A_a,X^i].  
\end{equation}
$C_n$ is a RR $n$-form and $i_X$ is the interior product with $X^i$
\begin{equation}
(i_X C_n)_{\mu_1\cdots \mu_{n-1}}=X^i (C_n)_{i\mu_1 \cdots \mu_{n-1}}.
\end{equation}
Also the second index in $\left(Q^{-1}-\delta\right)^{ij}$ is being raised using $E^{ij}$ which is the inverse of $E_{ij}$, that is $E^{ik}E_{kj}={\delta^i}_j$.
The determinants are over the spacetime indices, while the trace is over the matrix indices. Note that the D-brane action is gauge-invariant, so the action for fields satisfying the orbifold projection is the same as the original action. Thus, it is enough to restrict the form of the fields to the structure given in \eqref{eq:projections} at the end of the calculation. We will also use that the fields satisfying the orbifold projection are periodic in the $\mathbb{RP}^3$ orbit of the $S^3$. This implies that we can do the integral over the $\mathbb{RP}^3$ directions by doing it over the $S^3$ and dividing by two to take into account the difference in volume.

\subsection{Small amplitude expansion}

In order to completely determine the action, it is necessary to give an ordering prescription inside the trace. This has been shown to be a symmetrized trace up to quartic terms in the fields, but it is unknown for higher order terms \cite{Tseytlin:1997csa,Hashimoto:1997gm}, although there are concrete proposals for additional derivative contributions up to $O((\alpha')^4)$ \cite{Koerber:2002zb,Keurentjes:2004tu}. For this reason, and to simplify the calculation, we will do an expansion for small amplitudes and keep only the lowest terms. In the expansion we will introduce a small parameter $\epsilon\ll 1$:
\begin{equation}
\Phi^i=\epsilon \phi^i, \  \ A_M=\cA_M J+\epsilon a_M, \ \ A_A=\epsilon a_A.
\end{equation}
Where we have defined the matrix $J$ as
\begin{equation}
J=\sigma^3\otimes \mathbb{1}_{N_f}.
\end{equation}
We are denoting by $\cA_M$ the $\ell=0$ mode of the gauge potential and $a_M$ the remaining $\ell>0$ modes. We can identify $\cA_M$ as the gauge field dual to the current of the $U(1)_B$ symmetry. 

At leading order in the small $\epsilon$ expansion, the equations reduce to the (non-linear) equations for the $\cA_M$ gauge field. It follows from the gauge invariance of the D-brane action  that flat connections $\cF_{MN}=2\partial_{[M} \cA_{N]}=0$ satisfy the equations of motion to this order. Note that charged fields on the brane can source the field strength of the $\cA_M$ gauge field. However this will only happen at $O(\epsilon^2)$ in the equations of motion, since the charge current itself will be charge neutral. Therefore it is enough to keep only the quadratic terms in the field strength $\cF_{MN}$ in the action for the class of solutions we will study. Also up to $O(\epsilon^2)$ the action has only quadratic products of flavor matrices and thus the symmetrized trace is equivalent to the usual trace.

The action to next-to-leading order can then be split as\footnote{We are dropping a lower order contribution that depends on the embedding but is independent of the fields, so it does not affect to the discussion.}
\begin{equation}\label{eq:actionD71}
S_{D7}\simeq - T_7 \int d^4 x  d\Omega_3 d\rho \rho^3 \,{\rm Tr}_{2N_f}\left[ \cL_F+\cL_{kin}+\cL_m+\cL_{S^3} \right], 
\end{equation}
where
\begin{equation}
\begin{split}
\cL_F=& \frac{\lambda^2}{4} \cF^{MN} \cF_{MN},
\\[5pt]
\cL_{kin}=& \frac{\epsilon^2}{4}  \sqrt{\det G}  \Bigg[ G_{44}G^{MN} \Bigg( L\lambda \Big( i[a_M,J] \cD_N \phi^4+\cD_M \phi^4 i[a_N,J]\Big)+\lambda^2  \cD_M\phi^4 \cD_N\phi^4 \Bigg)
\\[5pt]
&+\lambda^2 G^{MN} G_{55} \cD_M\phi^5 \cD_N \phi^5 +\frac{1}{2} G^{MN}G^{PQ}  \hat{f}_{MP}\hat{f}_{NQ}+ G^{MN}G^{AB}  \hat{f}_{MA}\hat{f}_{NB}\Bigg],\\[5pt]
\cL_m=& \frac{\epsilon^2}{4}  \sqrt{\det G}  \Big[ L^2 \left( i[J,\phi^5]\right)^2 G_{44}G_{55} +G_{44}G^{ab} L^2 i[a_a,J] i[a_b,J]\Big],
\\[5pt]
\cL_{S^3}=& \frac{\epsilon^2}{4}  \sqrt{\det G}  \Big[ G_{44}G^{AB} \Big(L\lambda \big( i[a_A,J]\cD_B \phi^4+\cD_A \phi^4 i[a_B,J]\big)+\lambda^2 \cD_A\phi^4 \cD_B\phi^4\Big)
\\[5pt]
&+\lambda^2 G_{55}G^{AB} \cD_A\phi^5 \cD_B \phi^5+\frac{1}{2} G^{AD}G^{BC}  \hat{f}_{AB}\hat{f}_{DC}\Big]-\frac{\epsilon^2}{4} C_4  \epsilon^{ABC}  \hat{f}_{\rho A}\hat{f}_{BC} .
\end{split}
\end{equation}
Indices are raised with the induced metric, and we have defined the covariant derivatives (for any $X=\phi^i,a_M,a_A$) and the field strengths as
\begin{equation}
\cD_M X=\partial_M X+i [J,X] \cA_M, \ \ \hat{f}_{ab}=\lambda( \cD_a a_b-\cD_b a_a), \ \ \cA_A=0.
\end{equation}

\subsection{Integration over $S^3$}

Next, we expand in scalar ($\cY_S$) and vector ($\cY_V$) spherical harmonics as follows
\begin{equation}
\begin{split}
\cA_M=& \sqrt{2}\pi \cV_M \cY_S^0(\Omega_3),\\[5pt]
\phi^i=& \sum_{\ell \geq 0} \sum_{a=1}^{N_S^\ell} \sigma_{\ell}^{i\,a} (x) \cY_{S}^{\ell\,a}(\Omega_3), \\[5pt]  
 a_M=& \sum_{\ell \geq 1} \sum_{a=1}^{N_S^\ell} v_{\ell,M}^a(x) \cY_{S}^{\ell\, a}(\Omega_3), \\[5pt]
a_A=& \sum_{\ell \geq 1} \sum_{a=1}^{N_V^\ell} \eta_{\ell}^a(x) \cY_{V\,A}^{\ell\, a}(\Omega_3)+ \sum_{\ell \geq 1} \sum_{a=1}^{N_S^\ell} \tau_{\ell}^a(x) \nabla_A \cY_{S}^{\ell\, a}(\Omega_3).
\end{split}
\end{equation}
The number of modes for each $\ell$ is $N_S^\ell=(\ell+1)^2$ for scalar modes and $N_V^\ell=2\ell(\ell+1)$ for vector modes. 

The harmonics are eigenvectors of the $S^3$ Laplacian with eigenvalues (we follow the conventions of \cite{Kruczenski:2003be})
\begin{equation}\label{eq:sphharm1}
\nabla_{S^3}^2 \cY_S^\ell =-\ell(\ell+2) \cY_S^\ell , \ \ \nabla_{S^3}^2 \cY_{V\,A}^\ell -R_A^{\ B} \cY_{V\,B}^\ell=-(\ell+1)^2  \cY_{V\,A}^\ell. 
\end{equation}
In addition, the vector spherical harmonics satisfy
\begin{equation}\label{eq:sphharm2}
\nabla_A \cY_V^{\ell\, A} =0, \ \ \ \varepsilon_A^{\ BC}\nabla_B \cY_{V\,C}^{\ell,\pm} = \pm (\ell+1) \cY_{V\,A}^{\ell,\pm}.
\end{equation}
Note that $N_V^\ell$ is an even number, half of the harmonics correspond to the $\cY_{V\,A}^{\ell,+}$ and the other half to the $\cY_{V\,A}^{\ell,-}$. The harmonics form an orthonormal basis
\begin{equation}\label{eq:sphharm3}
\int d\Omega_3 \, \cY_S^{\ell\, a}\cY_S^{\ell'\, b}=\delta^{\ell\ell'}\delta^{ab}, \ \ \int d\Omega_3 \, g_{S^3}^{AB} \cY_{V\,A}^{\ell\, a}\cY_{V,B}^{\ell'\, b}=\delta^{\ell\ell'}\delta^{ab}.
\end{equation}

We introduce these expansions in \eqref{eq:actionD71}, integrate by parts and use \eqref{eq:sphharm1} and \eqref{eq:sphharm2} to get rid of derivatives along the $S^3$ directions and use \eqref{eq:sphharm3} to do the integrals. The orbifold plus orientifold projections constrain the fields to have the form in \eqref{eq:projected}. After integrating, the fields that are neutral under the $U(1)_B$ symmetry are completely decoupled from the rest at the order we are computing the action and we can turn them off. We will  reduce the action to a trace over $N_f\times N_f$ matrices using the results in Appendix~\ref{app:projaction}. The action becomes:  
\begin{equation}\label{eq:actionD72}
S_{D7}\simeq - T_7 \int d^4 x  d\rho \rho^3 \,{\rm Tr}_{N_f}\Bigg[ \cL_F+\sum_{\underset{  \text{odd}}{\small \ell\geq 1}} \sum_{a=1}^{N_S^\ell} \Big(\cL_\sigma^{\ell\, a}+\cL_\tau^{\ell\, a}+\cL_v^{\ell\, a}+\cL_{\rm mix}^{\ell\, a}\Big)+\sum_{\underset{  \text{even}}{\small \ell\geq 2}} \sum_{a=1}^{N_V^\ell/2} \sum_{\alpha=\pm} \cL_\eta^{\ell\, a\, \alpha} \Bigg], 
\end{equation}
where, omitting the $\ell$ and spherical harmonic indices,
\begin{equation}\label{eq:LYterms}
\begin{split}
\cL_F=& \frac{\pi^2\lambda^2}{2} \cV^{MN} \cV_{MN},
\\[5pt]
\cL_\sigma=&  \lambda^2\frac{\epsilon^2}{2}  \rho^3  \Bigg[ \frac{R^2}{r^2}G^{MN}( \cD_M\sigma^i)^\dagger \cD_N\sigma^i 
+ \frac{4L^2}{\lambda^2}   \frac{R^4}{r^4}(\sigma^5)^\dagger \sigma^5+\frac{\ell(\ell+2)}{\rho^2} (\sigma^i)^\dagger \sigma^i\Bigg],
\\[5pt]
\cL_\tau=&  \ell(\ell+2)\lambda^2\frac{\epsilon^2}{2}  \rho^3  \Bigg[  \frac{r^2}{R^2\rho^2}G^{MN}(\cD_M\tau)^\dagger \cD_N\tau
+\frac{4L^2}{\lambda^2}  \frac{1}{\rho^2} \tau^\dagger\tau\Bigg],
\\[5pt]
\cL_v=& \lambda^2\frac{\epsilon^2}{2}  \rho^3  \Bigg[ \Big( G^{MN}G^{PQ}-G^{MQ}G^{PN} \Big) (\cD_M v_P)^\dagger \cD_N v_Q+\ell(\ell+2) \frac{r^2}{R^2\rho^2} G^{MN}v_M^\dagger v_N + \frac{4L^2}{\lambda^2}  \frac{R^2}{r^2} G^{MN} v_M^\dagger v_N\Bigg],
\\[5pt]
\cL_\eta^\pm=&  \lambda^2\frac{\epsilon^2}{2}  \rho^3  \Bigg[ 
 \frac{r^2}{R^2\rho^2}G^{MN} (\cD_M\eta^\pm)^\dagger \cD_N\eta^\pm
+\frac{4L^2}{\lambda^2}  \frac{1}{\rho^2}(\eta^\pm)^\dagger \eta^\pm+ (\ell+1)^2 \frac{r^4}{R^4\rho^4}(\eta^\pm)^\dagger\eta^\pm 
\\[5pt]
& \mp(\ell+1)\frac{r^4}{\rho^3 R^4} \Big[\partial_\rho (\eta^\pm)^\dagger\eta^\pm+h.c. \Big] \Bigg],
\\[5pt]
\cL_{\rm mix}=&  \lambda^2\frac{\epsilon^2}{2}  \rho^3  \Bigg[ \frac{2i L}{\lambda} \frac{R^2}{r^2}G^{MN} \left( v_M^\dagger \cD_N \sigma^4-v_M (\cD_N \sigma^4)^\dagger\right)
-\ell(\ell+2) \frac{r^2}{R^2\rho^2} G^{MN}\Big[(\cD_M\tau)^\dagger v_N+h.c.\Big]
\\[5pt]
&+\frac{1}{\rho^2} \frac{2iL}{\lambda} \ell(\ell+2) \Big[\tau^\dagger\sigma^4-h.c. \Big] \Bigg] \, ,
\end{split}
\end{equation}
where the covariant derivative and field strength now are
\begin{equation}
\cD_M X= \partial_M X +2 i\cV_M X, \ \ \cV_{MN}=\partial_M\cV_N-\partial_N\cV_M.
\end{equation}
The charged fields are antisymmetric $N_f\times N_f$ matrices $X^T=-X$, while $\cV_M\propto \mathbb{1}_{N_f}$.

\section{Spontaneous breaking of baryon symmetry}\label{sec:density}

In the absence of light baryons, two different kind of phases with a non-zero density of baryon charge have been studied in holographic models with flavor branes. In one type, the baryon charge is sourced by strings extending between the flavor brane and the horizon. In the dual field theory this corresponds to a state where the charge is carried by quarks rather than baryons. In the other type of phase, there is a baryon vertex that sources the charge. The baryon vertex has equivalent descriptions as a brane wrapping the internal directions or a solitonic configuration on the brane with flux along the internal directions. This corresponds to a state with heavy baryons in the dual field theory.

In a theory of light baryons there is a third kind of state with non-zero charge density that can appear naturally. The light baryon operator, that is charged under baryon symmetry, can acquire a non-zero expectation value. In the gravity side, this corresponds to having a solution of finite energy density where the fields dual to the light baryon operators are turned on. Since they are charged under the $U(1)_B$ symmetry on the brane, these solutions will also support a non-zero flux.

In principle we expect that for a zero chemical potential the ground state has zero charge density. As the chemical potential is increased there can be a phase transition to the phase with spontaneously broken symmetry. A necessary condition is that there are solutions for the fields on the brane that satisfy the right boundary conditions, they have to be regular at the tip of the brane and normalizable close to the boundary. If such solutions exist, then one has to compare the free energy of each phase to determine which one is thermodynamically favored. In the following we will show that at large enough values of the chemical potential a phase with spontaneous symmetry breaking exists and is thermodynamically preferred over the zero charge phase.

\subsection{Small amplitude solutions of charged fields}

We work in the small amplitude expansion that we introduced in the previous section. To leading order, we just introduce a chemical potential $\mu$ as a constant value for the $U(1)_B$ gauge field 
\begin{equation}
\cV_M=\mu \delta_M^0.
\end{equation}
To the next order we have to solve the linear equations for the charged fields. For convenience we define
\begin{equation}
v_0=2i\mu \tilde{v}_0, \ \ \sigma^4=2i\frac{L}{\lambda}\tilde{\sigma}^4,
\end{equation}
and
\begin{equation}
\tilde{\mu}^2=4\left(\mu^2-\frac{L^2}{\lambda^2}\right).
\end{equation}
From \eqref{eq:actionD72} and \eqref{eq:LYterms} we get the following set of equations for each $\ell$ and spherical harmonic 
\begin{itemize}
\item $\sigma^5$
\begin{equation}\label{eq:s5}
0=\partial_\rho\left(\rho^3\partial_\rho \sigma^5\right)-\ell(\ell+2) \rho \sigma^5+\tilde{\mu}^2\rho^3\frac{R^4}{r^4}\sigma^5.
\end{equation}
\item $\eta$
\begin{equation}\label{eq:eta}
0=\partial_\rho\left(\rho\frac{r^4}{R^4}\partial_\rho \eta^\pm\right)-(\ell+1)^2 \frac{r^4}{\rho R^4} \eta^\pm+\tilde{\mu}^2\rho\eta^\pm \mp (\ell+1)\partial_\rho\left(\frac{r^4}{R^4}\right)\eta^\pm.
\end{equation}
\item $v_i$
\begin{equation}\label{eq:vi}
0=\partial_\rho(\rho^3 \partial_\rho v_i)-\ell(\ell+2)\rho v_i+\tilde{\mu}^2 \rho^3 \frac{R^4}{r^4}v_i.
\end{equation}
\item $\tilde{v}_0$
\begin{equation}\label{eq:v0}
0=\partial_\rho(\rho^3(v_\rho-\partial_\rho\tilde{v}_0))+\ell(\ell+2) \rho (\tilde{v}_0-\tau)+\frac{4L^2}{\lambda^2}\rho^3\frac{R^4}{r^4}(\tilde{v}_0-\tilde{\sigma}^4).
\end{equation}
\item $v_\rho$
\begin{equation}\label{eq:vr}
0=\ell(\ell+2)\frac{r^4}{R^4\rho^2} (v_\rho-\partial_\rho\tau)-4\mu^2(v_\rho-\partial_\rho\tilde{v}_0)-\frac{4L^2}{\lambda^2}(\partial_\rho \tilde{\sigma}^4-v_\rho).
\end{equation}
\item $\tau$
\begin{equation}\label{eq:t}
0=\partial_\rho\left[ \rho\frac{r^4}{R^4}(v_\rho-\partial_\rho\tau)\right]+4\mu^2\rho(\tilde{v}_0-\tau)-\frac{4L^2}{\lambda^2}\rho^3(\tilde{\sigma}^4-\tau).
\end{equation}
\item $\sigma^4$
\begin{equation}\label{eq:s4}
0=\partial_\rho(\rho^3(\partial_\rho\tilde{\sigma}^4-\tilde{v}_\rho))-\ell(\ell+2)\rho(\tilde{\sigma}^4-\tau)+4\mu^2\rho^3\frac{R^4}{r^4}(\tilde{\sigma}^4-\tilde{v}_0).
\end{equation}
\end{itemize}
Note that there is a residual gauge symmetry that leaves the equations invariant
\begin{equation}
\delta\tilde{v}_0= \delta\tau=\delta\tilde{\sigma}^4=\alpha, \ \ \delta v_\rho=\partial_\rho\alpha.
\end{equation}
We can then fix the gauge to $\tilde{v}_0=0$. If we add \eqref{eq:v0}+\eqref{eq:s4}, we get a decoupled equation for $\tilde{\sigma}^4$ which is the same as the equation for $\sigma^5$ \eqref{eq:s5}
\begin{equation}\label{eq:s4d}
0=\partial_\rho\Big(\rho^3\partial_\rho\tilde{\sigma}^4\Big)-\ell(\ell+2)\rho\tilde{\sigma}^4+\tilde{\mu}^2\rho^3\frac{R^4}{r^4} \tilde{\sigma}^4.
\end{equation}
On the other hand, we can check that the equations are not all independent, since $\partial_\rho( \rho^3$ \eqref{eq:vr}$)$ $-\ell(\ell+2)$\eqref{eq:t} $+4\mu^2$ \eqref{eq:v0} $+\frac{4L^2}{\lambda^2}$\eqref{eq:s4} $=0$. We will discard \eqref{eq:t}. From \eqref{eq:v0} we get
\begin{equation}
\tau=\frac{1}{\ell(\ell+2)} \left[ \frac{1}{\rho}\partial_\rho \Big(\rho^3v_\rho \Big)-\frac{4L^2}{\lambda^2} \rho^2\frac{R^4}{r^4}\tilde{\sigma}^4\right].
\end{equation}
Substituting in \eqref{eq:vr}, we get
\begin{equation}\label{eq:vr2}
0=\partial_\rho\left[\frac{1}{\rho}\partial_\rho \Big(\rho^3v_\rho \Big) \right]-\ell(\ell+2)v_\rho+\tilde{\mu}^2\rho^2\frac{R^4}{r^4}v_\rho-\frac{4L^2}{\lambda^2} \partial_\rho\left( \rho^2\frac{R^4}{r^4}\right)\tilde{\sigma}^4.
\end{equation}

Except for $\tau$ and $v_\rho$, the equations of motion are the same as for the fluctuations in  \cite{Kruczenski:2003be}, with the substitution of the mass by the chemical potential $M^2\longrightarrow \tilde{\mu}^2$. The relation is 
\begin{equation}
\sigma^5, \tilde{\sigma}^4 \to \phi, \ \ \eta^\pm\to\phi_I^\pm, \ \ v_i\to\phi_{II}, \ \ v_\rho\to \phi_{III},\ \ \tau\to \tilde{\phi}_{III}. 
\end{equation}
For $v_\rho$ and $\tau$ the difference are the inhomogeneous terms proportional to $\tilde{\sigma}^4$. The analysis in \cite{Kruczenski:2003be} showed that there are solutions that are both regular and normalizable for discrete values of the mass. Since the equations we have derived are almost identical, we expect that in our case for a set of discrete values of the chemical potential regular and normalizable solutions to exist. We only need to determine the effect of the additional terms in the equation for $v_\rho$. Let us denote by $v_\rho^{(h)}$ the solution to the homogeneous equation, which is the same as in  \cite{Kruczenski:2003be}, and write
\begin{equation}
v_\rho=v_\rho^{(h)} V.
\end{equation}
The expansion of the solutions close to the boundary and at the tip of the brane are
\begin{equation}
\begin{split}
&v_\rho^{(h)} \underset{ \rho\to\infty}{ \sim} \frac{1}{\rho^{\ell+3}},\ \ v_\rho^{(h)} \underset{\rho\to 0}{\sim} \rho^{\ell-1},\\[5pt]
&\tilde{\sigma}^4 \underset{ \rho\to\infty}{ \sim} \frac{1}{\rho^{\ell+2}},\ \ \tilde{\sigma}^4 \underset{\rho\to 0}{\sim} \rho^{\ell} \, .
\end{split}
\end{equation}
Then \eqref{eq:vr2} becomes
\begin{equation}
0=\partial_\rho^2 V+\left(\frac{5}{\rho}+2\frac{\partial_\rho v_\rho^{(h)}}{v_\rho^{(h)}}  \right)\partial_\rho V-\frac{4L^2}{\lambda^2} \partial_\rho\left( \rho^2\frac{R^4}{r^4}\right)\frac{\tilde{\sigma}^4}{v_\rho^{(h)}}.
\end{equation}
Let us define
\begin{equation}
K=\rho^5 (v_\rho^{(h)})^2,
\end{equation}
so the coefficient of $\partial_\rho V$ is $\partial_\rho K/K$. Then, the solution for the first derivative is
\begin{equation}
\partial_\rho V=\frac{4L^2}{\lambda^2} \frac{1}{K}\int_0^\rho K \partial_{\tilde{\rho}}\left( \tilde{\rho}^2\frac{R^4}{r^4}\right)\frac{\tilde{\sigma}^4}{v_\rho^{(h)}}.
\end{equation}
The limit of the integral is fixed by regularity of the solution at $\rho=0$. Close to the boundary
\begin{equation}
\partial_\rho V\underset{\rho\to\infty}{\sim} \rho^{2\ell+1}.
\end{equation}
Therefore,
\begin{equation}
v_\rho \underset{\rho\to\infty}{\sim} \rho^{\ell-1}
\end{equation}
Since this turns on a non-normalizable contribution, we would have to add to $v_\rho$ a non-normalizable homogeneous solution such that the leading terms are canceled out. However, for the values of $\tilde{\mu}$ for which there is a normalizable and regular $\tilde{\sigma}^4$, the non-normalizable homogeneous solution of $v_\rho$ is not regular. Then, we have to set $\tilde{\sigma}^4=0$. 

With this condition the equations are the same as in \cite{Kruczenski:2003be} , so we can borrow their results for the meson spectrum with the proper identification between the mass and the chemical potential. The critical values of the chemical potential for which regular and normalizable solutions exist are
\begin{itemize}
\item $\sigma^5$
\begin{equation}\label{eq:mus5}
\tilde{\mu}^2 =\frac{4L^2}{R^4} (n+\ell+1)(n+\ell+2), \ \ n=0,1,2,\dots, \ \ \ell\geq 1\ {\rm odd}.
\end{equation}
\item $\eta$
\begin{equation}
\begin{split}
(\tilde{\mu}^{+})^2=& \frac{4L^2}{R^4} (n+\ell+2)(n+\ell+3), \ \ n=0,1,2,\dots,\ \ \ell\geq 2 \ {\rm even},\\[5pt]
(\tilde{\mu}^{-})^2=& \frac{4L^2}{R^4} (n+\ell)(n+\ell+1), \ \ n=0,1,2,\dots,\ \ \ell\geq 2 \ {\rm even}.
\end{split}
\end{equation}
\item $v_i$
\begin{equation}
\tilde{\mu}^2=\frac{4L^2}{R^4} (n+\ell+1)(n+\ell+2), \ \ n=0,1,2,\dots,\ \ \ell\geq 1\ {\rm odd}.
\end{equation}
\item $v_\rho$, $\tau$
\begin{equation}
\tilde{\mu}^2=\frac{4L^2}{R^4} (n+\ell+1)(n+\ell+2), \ \ n=0,1,2,\dots,\ \ \ell\geq 1\ {\rm odd}.
\end{equation}
\end{itemize}
The lowest value is for $n=0$ and $\ell=1$ ($\sigma^5$, $v_i$, $v_\rho$, $\tau$) or $\ell=2$ ($\eta^-$). This gives the critical chemical potential
\begin{equation}\label{eq:muc}
\mu_c^2=\frac{L^2}{\lambda^2}+\frac{6L^2}{R^4}=\frac{R^4}{\lambda^2}\left( \frac{L^2}{R^4}+\frac{\lambda^2}{R^4}\frac{6L^2}{R^4} \right).
\end{equation}
Note that $R^4/\lambda^2=\frac{\lambda_{YM}}{(2\pi)^2}$, where $\lambda_{YM}$ is the 't Hooft coupling in the dual field theory. On the other hand, $L/R^2$ is  the mass scale of mesons, the mass gap is $m_{gap}^2=8L^2/R^4$ \cite{Kruczenski:2003be}. Then, 
\begin{equation}
\mu_c^2=\frac{\lambda_{YM}}{8(2\pi)^2}\left(1+\frac{6(2\pi)^2}{\lambda_{YM}} \right) m_{gap}^2.
\end{equation}
This implies that $\mu_c$ is of the order of the quark mass $m_q\sim \sqrt{\lambda_{YM}}m_{gap}$.

\subsection{Backreaction on the gauge field}

We have found that when $\mu=\mu_c$, there are regular and normalizable solutions for $\sigma_{\ell=1}^5$ (4 modes), $\eta_{\ell=2}^{-}$ (6 modes), $v_{\ell=1,\, i}$ (4 modes) and $v_{\ell=1,\, \rho}$, $\tau_{\ell=1}$ (4 modes). These are  members of the lowest supermultiplet in the antisymmetric representation ($j_L=1/2$). If any (possibly all) of them are non-zero, there will be expectation values for charged operators in the dual field theory. The solution will then be dual to a superfluid state with spontaneously broken baryon symmetry. If $v_{\ell=1,\, i}\neq 0$ the condensate will have a $p$-wave component  and rotational invariance will be broken as well. In contrast to the usual D3/D7 model here scalar condensation can happen at the same time as vector condensation.

Let us compute the equation of motion for the gauge field. Taking a variation of the small amplitude action \eqref{eq:actionD72}, \eqref{eq:LYterms} with respect to the gauge potential we find
\begin{equation}
\partial_N \left(\sqrt{-G} \cV^{NM} \right)=\frac{\epsilon^2}{2\pi^2}  J^M.
\end{equation}
We can split the current as
\begin{equation}
J^M=\frac{1}{N_f}{\rm Tr}_{N_f}\Bigg[ \sum_{\underset{  \text{odd}}{\small \ell\geq 1}} \sum_{a=1}^{N_S^\ell}  \Big( J^{M\,\ell\, a}_{\sigma}+J^{M\,\ell\, a}_{v}+J^{M\,\ell\, a}_\tau \Big)+ \sum_{\underset{  \text{even}}{\small \ell\geq 2}} \sum_{a=1}^{N_V^\ell/2} \sum_{\alpha=\pm} J^{M\,\ell\, a\,\alpha}_{\eta}\Bigg],
\end{equation}
where, omitting the $\ell$ and spherical harmonic indices, 
\begin{equation}
\begin{split}
J^M_{\sigma}= & i \rho^3G^{MN}\frac{R^2}{r^2} \Big[ \big( \cD_N\sigma^5 \big)^\dagger \sigma^5-(\sigma^5)^\dagger \cD_N\sigma^5 \Big],
\\[5pt]
J^M_v=& i \rho^3 \Big(G^{MN}G^{PQ}-G^{MQ}G^{NP} \Big) \Big[ \big( \cD_N v_Q \big)^\dagger v_P- v_P^\dagger \cD_N v_Q\Big],
\\[5pt]
J^M_\tau=& i \rho G^{MN} \frac{r^2}{R^2}\ell(\ell+2) \Big[ \big( \cD_N\tau-v_N \big)^\dagger \tau-\tau^\dagger \big( \cD_N\tau-v_N \big) \Big],
\\[5pt]
J^{M\,\pm}_\eta=&  i\rho G^{MN}\frac{r^2}{R^2} \Big[ \big( \cD_N\eta^\pm \big)^\dagger \eta^ \pm-(\eta^\pm)^\dagger \cD_N\eta^\pm \Big].
\end{split}
\end{equation}
Since the charged fields are antisymmetric in flavor indices, they will not source the non-Abelian components of the gauge field if $N_f=2$. For $N_f>2$ they can be ignored at this level of the small amplitude analysis, as we have the freedom to select any configuration of the charged fields, and in particular pick one that only sources the Abelian gauge field. A full analysis taking into account the non-Abelian components is beyond the scope of this paper and we will postpone it for future work.

Taking into account that the solutions for homogeneous fields are time-independent and homogeneous, and that the dependence on $\rho$ is the same for a field and its hermitian conjugate (only an overall constant coefficient changes), one can show that $J^\rho=0$. Therefore, the field strength is time-independent and homogeneous as well
\begin{equation}
\partial_\rho \Big( \rho^3 \eta^{\mu\nu}\cV_{\rho \nu} \Big)=\frac{\epsilon^2}{2\pi^2} J^\mu.
\end{equation}
Where
\begin{equation}
\begin{split}
J^0_\sigma= & -4\mu \rho^3\frac{R^4}{r^4} (\sigma^5)^\dagger \sigma^5,\\[5pt]
J^0_v=& -4\mu \rho^3 \left( \frac{R^4}{r^4} \delta^{ij}v_i^\dagger v_j+v_\rho^\dagger v_\rho \right),\\[5pt]
J^0_\tau=& -4\mu \rho \ell(\ell+2)\tau^\dagger \tau,\\[5pt]
J^{0\,\pm}_\eta=&  -4\mu  \rho (\eta^\pm)^\dagger \eta^\pm.
\end{split}
\end{equation}
and
\begin{equation}
\begin{split}
J^i_\sigma= & 0,\\[5pt]
J^i_v=& i \rho^3\delta^{ij} \left(v_\rho^\dagger \partial_\rho v_j-\partial_\rho v_j^\dagger v_\rho  \right),\\[5pt]
J^i_\tau=& i \rho\delta^{ij} \ell(\ell+2)\left(\tau^\dagger v_j-v_j^\dagger \tau\right),\\[5pt]
J^{i\,\pm}_\eta=& 0.
\end{split}
\end{equation}
Using the equation of motion for $\tau$:
\begin{equation}
\tau=\frac{1}{\ell(\ell+2)}\frac{1}{\rho}\partial_\rho \Big( \rho^3 v_\rho \Big),
\end{equation}
the spatial components of the current becomes a total derivative
\begin{equation}
J^i_v+J^i_\tau=i\delta^{ij} \partial_\rho \Big[\rho^3 \Big( v_\rho^\dagger  v_j- v_j^\dagger v_\rho \Big) \Big].
\end{equation}
This leads to the equation
\begin{equation}
\cV_{\rho i}=\frac{1}{2\pi^2}\sum_{\underset{  \text{odd}}{\small \ell\geq 1}} \sum_{a=1}^{N_S^\ell} \frac{1}{N_f}{\rm Tr}_{N_f} \Big[i\epsilon^2 \Big( v_\rho^\dagger  v_i- v_i^\dagger v_\rho \Big) \Big].
\end{equation}
The normalizable solution would be
\begin{equation}
\cV_i=\frac{1}{2\pi^2}\sum_{\underset{  \text{odd}}{\small \ell\geq 1}} \sum_{a=1}^{N_S^\ell}\frac{1}{N_f}{\rm Tr}_{N_f}\Big[-i\epsilon^2\int^\infty_\rho \Big( v_\rho^\dagger  v_i- v_i^\dagger v_\rho \Big)\Big].
\end{equation}
Note that this contribution vanishes for $N_f=2$, or in some other cases, for instance if $v_i$ and $v_\rho$ are proportional to the same flavor matrix.

For the time component, the regular solutions are
\begin{equation}
\partial_\rho \cV_0=-\frac{1}{2\pi^2}\frac{\epsilon^2}{\rho^3}\int_0^\rho J^0,
\end{equation}
and the normalizable solution is
\begin{equation}
\cV_0=\frac{1}{2\pi^2}\epsilon^2\int_\rho^\infty \frac{d\rho_1}{\rho_1^3}\int_0^{\rho_1} J^0.
\end{equation}

\subsection{Free energy}

According to the usual AdS/CFT dictionary, the free energy density is equal to minus the on-shell action density (in Lorentzian signature)
\begin{equation}
\cF=-\cL_{D7}\simeq T_7 \int   d\rho \rho^3 \,{\rm Tr}_{N_f} \Bigg[ \cL_F+\sum_{\underset{  \text{odd}}{\small \ell\geq 1}} \sum_{a=1}^{N_S^\ell} \Big( \cL_\sigma^{\ell\, a}+\cL_\tau^{\ell\, a}+\cL_v^{\ell\, a}+\cL_{\rm mix}^{\ell\, a} \Big)+\sum_{\underset{  \text{even}}{\small \ell\geq 2}} \sum_{a=1}^{N_V^\ell/2} \sum_{\alpha=\pm} \cL_\eta^{\ell\, a\, \alpha} \Bigg],
\end{equation}
where each of the terms \eqref{eq:LYterms} has to be evaluated on the solutions to the equations of motion. Integrating by parts the terms with derivatives acting on the $(\sigma^i)^\dagger, v_M^\dagger, \tau^\dagger$ and $(\eta^\pm)^\dagger$ and using the equations of motion for the fields $\sigma^i, v_M, \tau$ and $\eta^\pm$, the action for the charged fields reduces to a boundary term, which for each $\ell$ and spherical harmonic component has the form
\begin{equation}
\begin{split}
\cL_{\rm charged }\sim & \lim_{\rho\to \infty} \lambda^2\frac{\epsilon^2}{2}\rho^3 \Bigg[(\sigma^5)^\dagger\partial_\rho \sigma^5+\delta^{ij} v_i^\dagger \partial_\rho v_j+\ell_\tau(\ell_\tau+2)\frac{r^4}{R^4\rho^2}\tau^\dagger \Big( \partial_\rho\tau-v_\rho \Big)
\\[5pt]
&+\sum_{\alpha=\pm} \frac{r^4}{R^4 \rho^2}(\eta^\alpha)^\dagger\partial_\rho\eta^\alpha+(\ell_\eta+1)\frac{r^4}{R^4\rho^3} (\eta^\alpha)^\dagger\eta^\alpha \Bigg].
\end{split}
\end{equation}
However, the solutions we are considering are normalizable, so they vanish as $\rho\to \infty$ in such a way that the contribution to the on-shell action vanishes as well $\cL_{\rm charged}=0$. 

Then, the free energy density is determined by the action for the gauge field
\begin{equation}
\cF\simeq \frac{N_f}{2} \pi^2\lambda^2 T_7 \int_0^{\infty}   d\rho \rho^3\, \cV^{MN}\cV_{MN} =- N_f \pi^2\lambda^2 T_7 \int_0^\infty   d\rho \rho^3  \Big[ \big( \partial_\rho \cV_0 \big)^2 - \big( \partial_\rho\cV_i \big)^2 \Big].
\end{equation}
The ground state of the system is the one of lowest free energy. If the charge is zero, then $\partial_\rho\cV_M=0$ and the free energy density vanishes. From the expression above we see that when the charged fields are non-zero the free energy becomes negative generically, in particular if $\cV_i=0$. Therefore, states with non-zero charge density are thermodynamically favored and as soon as the chemical potential reaches the critical value, a condensate will form and the $U(1)_B$ symmetry will be spontaneously broken. However, at this order in the expansion the amplitude of the charged fields can have arbitrary values (as long as the expansion does not break down), so we cannot determine the endpoint of the condensation.

\section{Ground state in a simple case}\label{sec:ground}

If the amplitude of the fields becomes large one can run into the problem of not having a full definition of the non-Abelian brane action. This could be avoided when the action has an expansion in $\lambda$ such that one can consistently keep the lower order terms. The usual flat spacetime expansion of \cite{Myers:1999ps} in the gauge field $A_a$ and scalars $\Phi^i$ does not work in this case because of the additional terms that appear when there are two stacks of probe branes separated by a distance $2L$.  We can nevertheless attempt an expansion in $\lambda$ by taking $L\sim \lambda$
\begin{equation}
L=\frac{\lambda}{R^2}\tilde{L}, \ \ \tilde{L}\sim R.
\end{equation}
Physically, this choice corresponds to making the mass of the quarks of order one in the strong coupling expansion $m_q\sim O(\lambda_{YM}^0)$.

Even with the expansion above, the phase diagram of the model with light baryons that we are studying is potentially quite complicated, since there are many charged operators that can condense as soon as the chemical potential reaches the critical value. However, in order to show the existence of a charged ground state, it would be enough to introduce a consistent ansatz such that only a few fields on the brane are turned on. Thus, we will make a series of simplifications that will allow us to reduce to a simple system corresponding to a particular section of the phase diagram where only one type of scalar operator condenses.

\subsection{Effective action and equations}

We saw that for small amplitudes $\Phi^4=0$, so we will take this condition to be part of the ansatz. This simplifies the matrix $Q$
\begin{equation}
{Q^4}_4= {Q^5}_5=\mathbb{1}, \ \ {Q^4}_5=L G_{55} \Big( i[J,\Phi^5] \Big) ,\ \   {Q^5}_4=L G_{44} \Big( i[\Phi^5,J] \Big).
\end{equation}
Such that
\begin{equation}
\det Q=\mathbb{1}+ L^2 \Big( i[J,\Phi^5] \Big)^2 G_{44}G_{55}.
\end{equation}
The main complication of the action is the non-quadratic coupling between charged fields. In order to gain some understanding of the charged phase we will further simplify our action by setting the charged vector fields to zero, so we are left with $\Phi^5$ and the $U(1)_B$ gauge field. In this case
\begin{equation}
P \Big[ E_{ai} \Big( Q^{-1}-\delta \Big)^{ij}E_{jb} \Big]=0.
\end{equation}
The Born-Infeld action is
\begin{equation}
\begin{split}
S_{BI}=& -\frac{1}{2} T_7 \int d^4 x  d\Omega_3 d\rho \,{\rm Tr}_{2N_f} \cL_{BI},\\[5pt]
\cL_{BI}=&\rho^3\sqrt{\mathbb{1}_{2N_f}+L^2 G_{44}G_{55} \Big( i[J,\Phi^5] \Big)^2}\sqrt{\det \Big[\delta_a^{\ b}\mathbb{1}_{2 N_f} +\lambda^2 G_{55}G^{bc}D_a \Phi^5 D_c \Phi^5  +\lambda G^{bc}F_{ac} \Big]}. 
\end{split}
\end{equation}
We will impose that the solutions are static and homogeneous in the field theory directions, so the only dependence of the fields is along the $S^3$ directions and the radial coordinate, and we will turn on only the time  component of the $U(1)_B$ gauge field.

We will also restrict to the simpler case of $N_f=2$, for which the charged fields are $SU(2)$ singlets, so we can neglect the non-Abelian part of the flavor gauge field. We will also turn off the even modes that are not charged under the baryon symmetry after the projection (but charged under the $SU(2)$ flavor symmetry). The two sectors could be coupled only by quartic terms, so there are always solutions where the $U(1)_B$ neutral fields are turned off. However, the coupling between the two sectors may trigger an instability such that the $U(1)_B$ neutral fields become nonzero, we will not consider this possibility here.

For $N_f=2$ ($\sigma^2$ is the antisymmetric Pauli matrix) 
\begin{equation}
\Phi^5=\left( \begin{array}{cc}
0 & \phi^5 \sigma^2 \\
\overline{\phi^5} \sigma^2 & 0
\end{array}
\right).
\end{equation}
Note that
\begin{equation}
\Big( i[J,\Phi^5] \Big)^2=4 \left|\phi^5\right|^2\mathbb{1}_{4}.
\end{equation}
Expanding to $O(\lambda^2)$
\begin{equation}
\begin{split}
S_{D7}\simeq & - 2T_7 \int d^4 x  d\Omega_3 d\rho \, \cL_{D7},
\\[5pt]
\cL_{D7}=& \rho^3 \Bigg[ 1+2L^2 G_{44}G_{55}|\phi^5|^2+2\lambda^2G_{55}G^{00}\cA_0^2|\phi^5|^2 
+\frac{\lambda^2}{2} G_{55} \Big( G^{\rho\rho}|\partial_\rho\phi^5|^2 +G^{AB} \partial_A\overline{\phi^5}\partial_B\phi^5 \Big)
\\[5pt]
& +\frac{\lambda^2}{2} G^{\rho\rho}G^{00} \Big(\partial_\rho \cA_0 \Big)^2 \Bigg].
\end{split}
\end{equation}
Expanding in spherical harmonics
\begin{equation}
\phi^5=\sum_{\underset{odd}{\small \ell \geq 1}} \sum_{a=1}^{N_S^\ell} \sigma_\ell^a \cY_S^{\ell\,a} (\Omega_3),\ \ \cA_0=\sqrt{2}\pi\cV_0\cY_S^0(\Omega_3),
\end{equation}
and  integrating over the $S^3 $ directions (we omit the $\ell$ and spherical harmonic indices in the second line)
\begin{equation}
\begin{split}
S_{D7}\simeq & - 2T_7 \int d^4 x  d\rho \, \Bigg[ \rho^3+\cL_\cV+\sum_{\underset{odd}{\small \ell \geq 1}} \sum_{a=1}^{N_S^\ell}   \cL_{\sigma}^{\ell\, a} \Bigg],
\\[5pt]
\cL_{\sigma}=& \rho^3 \Bigg[ \frac{2L^2R^4}{(\rho^2+L^2)^2}|\sigma|^2-2\lambda^2\frac{R^4}{(\rho^2+L^2)^2}\cV_0^2|\sigma|^2 
+\frac{\lambda^2}{2} \Big( |\partial_\rho\sigma|^2 +\frac{\ell(\ell+2)}{\rho^2} |\sigma|^2 \Big) \Bigg],
\\[5pt]
\cL_\cV=& -\pi^2\lambda^2 \rho^3  \Big( \partial_\rho \cV_0 \Big)^2.
\end{split}
\end{equation}
The equations of motion are
\begin{equation}
\begin{split}
&0=\sigma''+\frac{3}{\rho}\sigma' -\frac{\ell(\ell+2)}{\rho^2} \sigma+\frac{4R^4}{(\rho^2+L^2)^2}\left( \cV_0^2-\frac{L^2}{\lambda^2}\right)\sigma ,\\[5pt]
&0=\cV_0''+\frac{3}{\rho}\cV_0'-\frac{2R^4}{\pi^2 (\rho^2+L^2)^2} |\sigma|^2\cV_0 .
\end{split}
\end{equation}
We can eliminate the dependence on $\lambda$, $L$ and $R$ from the equations through the following change of variables
\begin{equation}
\rho=L x, \ \ \cV_0=\frac{L}{\lambda}+\frac{\lambda L}{4R^4}\alpha_0,\ \ \sigma=\pi\frac{\lambda L}{2\sqrt{2} R^4} s.
\end{equation}
Then, to leading order in the $\lambda$ expansion, the equations become
\begin{equation}\label{eq:numer}
\begin{split}
&0=s''+\frac{3}{x}s'-\frac{\ell(\ell+2)}{x^2}s+\frac{2\alpha_0}{(x^2+1)^2}s ,\\[5pt]
&0=\alpha_0''+\frac{3}{x}\alpha_0'-\frac{|s|^2}{(x^2+1)^2} .
\end{split}
\end{equation}
These resemble the typical equations of charged holographic superconductors in the probe approximation. A difference is that the effective background geometry has no horizon, but ends smoothly at $x=0$.

\subsection{Solutions}

The equations \eqref{eq:numer} are non-linear and have no known analytical solutions. We resort to numerics to find a solution for the lowest angular momentum $\ell=1$. First, we change coordinates to constraint the calculation to a finite interval. We define
\begin{equation}
 s(x)=x \,y(x),\ \ x=\frac{1}{u}-1,
\end{equation}
so that the boundary is at $u=0$ and the tip of the brane at $u=1$. Normalizable solutions have the following expansions at the boundary
\begin{equation}
y=\sum_{n\geq 4} y_B^{(n)} u^n, \ \ \alpha_0=\alpha_B^{(0)}+\sum_{n \geq 2} \alpha_B^{(n)}  u^n \, ,
\end{equation}
with undetermined coefficients $y_B^{(4)}$, $\alpha_B^{(0)}$ and $\alpha_B^{(2)}$. \\
Regular solutions at the tip of the brane have an expansion
\begin{equation}
y=y_H^{(0)}+  \sum_{n\geq 2} y_H^{(n)} (u-1)^n, \ \ \alpha_0=\alpha_H^{(0)}+\sum_{n \geq 4}  \alpha_H^{(n)} (u-1)^n.
\end{equation}
The undetermined coefficients are $y_H^{(0)}$ and $\alpha_H^{(0)}$.

\begin{figure}[t]
\centering
\includegraphics[width=7cm]{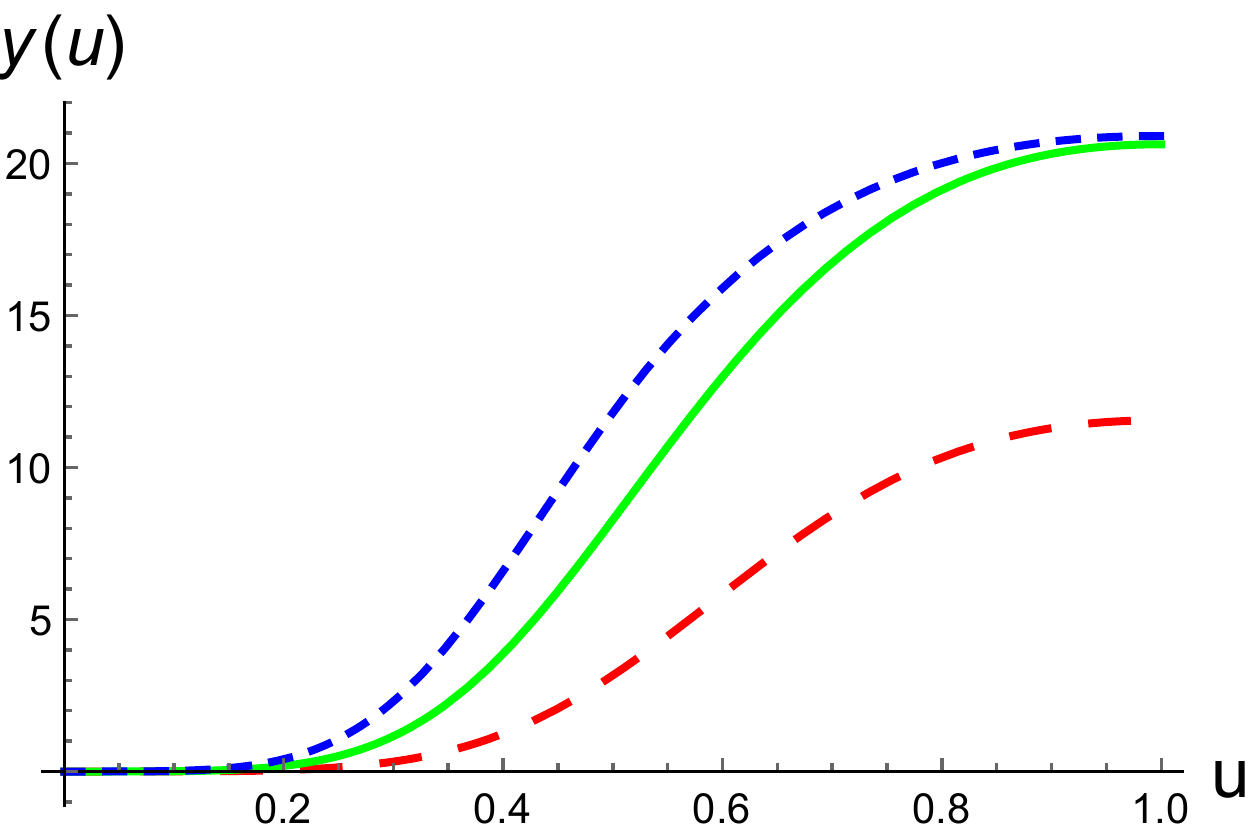}
\quad
\includegraphics[width=7cm]{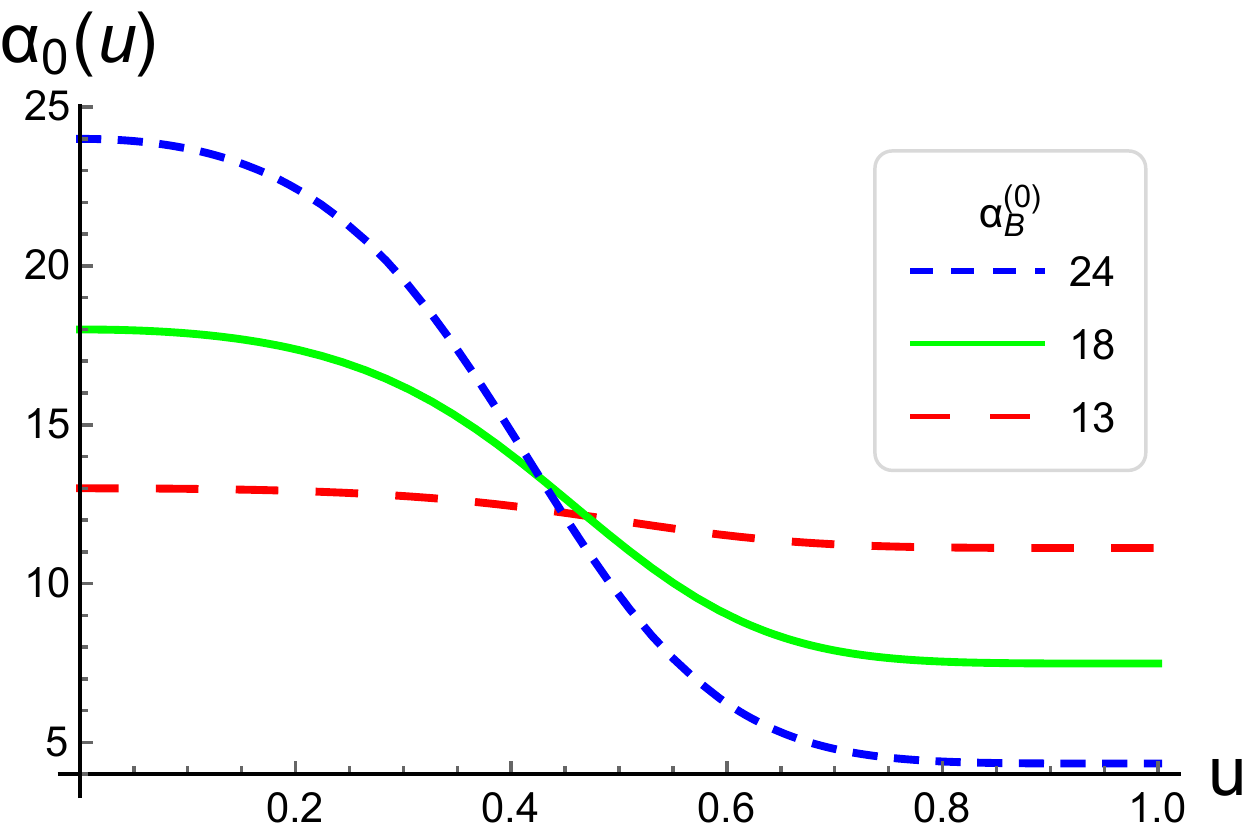}
\caption{
Scalar (left) and gauge field (right) solutions as functions of the radial coordinate for three different values of  $\alpha_B^{(0)}$. The boundary is at $u=0$ and the tip of the brane at $u=1$.
}
\label{flow}
\end{figure}

The coefficient $\alpha_B^{(0)}$ determines the value of the chemical potential and we will take its value as an input for our calculation. To find the solution we shoot from the boundary ($u=0$) and from the tip ($u=1$) of the brane and match the two solutions at $u=1/2$. There are four matching conditions from taking the values of $y$ and $\alpha_0$ and of their first derivatives to be the same at the midpoint. These four conditions fix the value of the coefficients $y_B^{(4)}$, $\alpha_B^{(2)}$, $y_H^{(0)}$ and $\alpha_H^{(0)}$. In Figure~\ref{flow} we plot some of the numerical solutions we find with this method.

%
%

\begin{figure}[t]
\centering
\includegraphics[width=7cm]{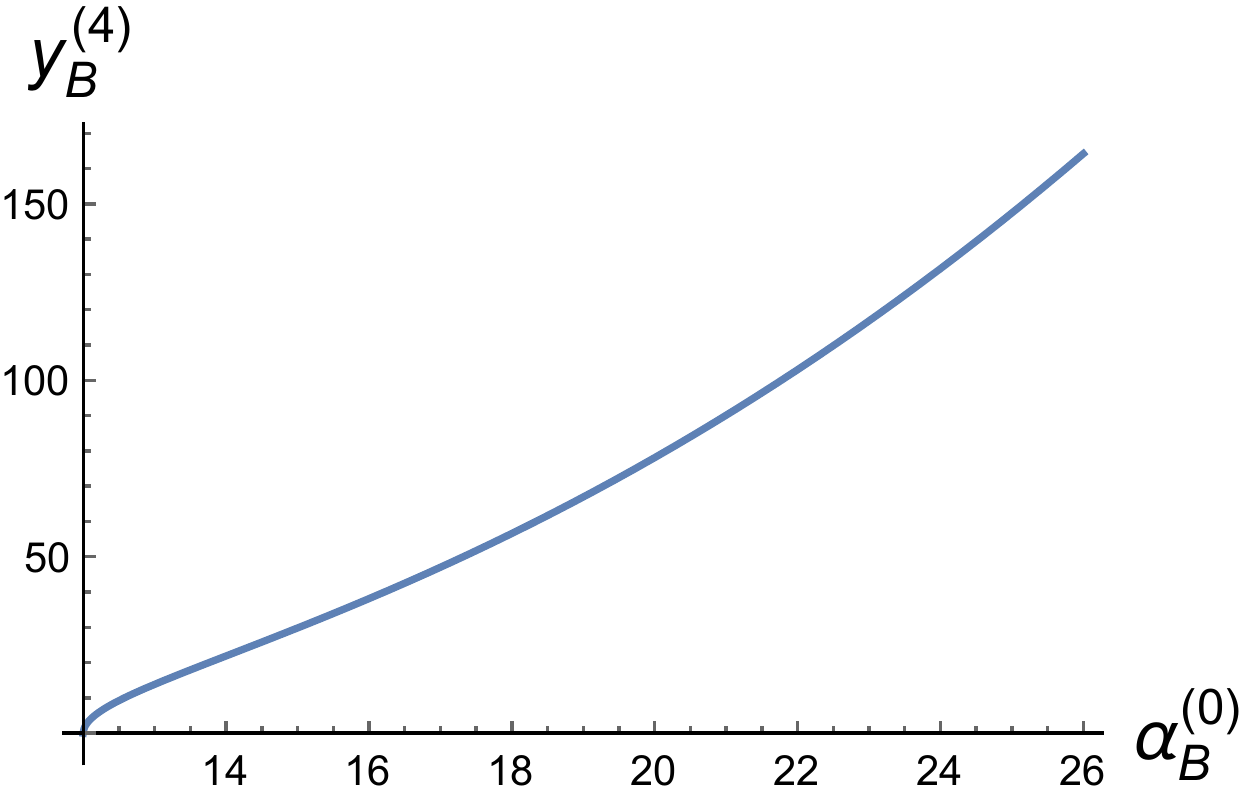}
\quad
\includegraphics[width=7cm]{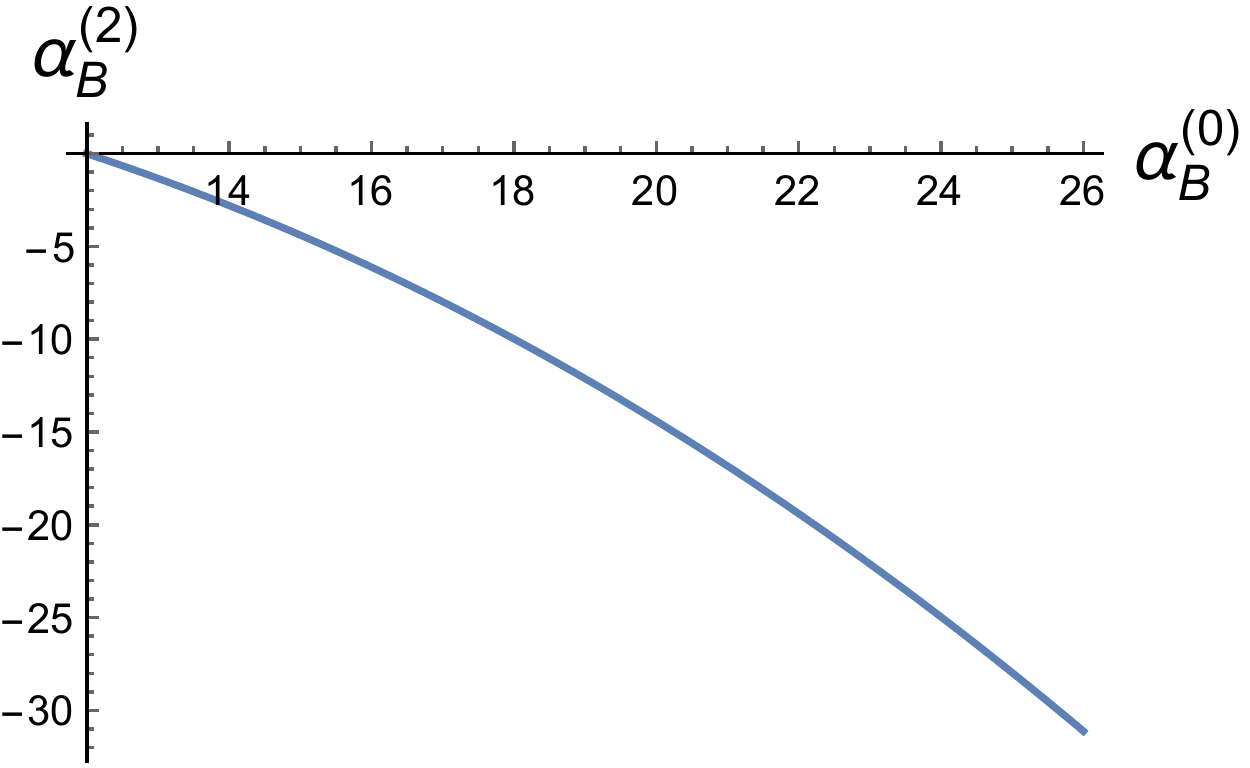}\\[10pt]
\includegraphics[width=7cm]{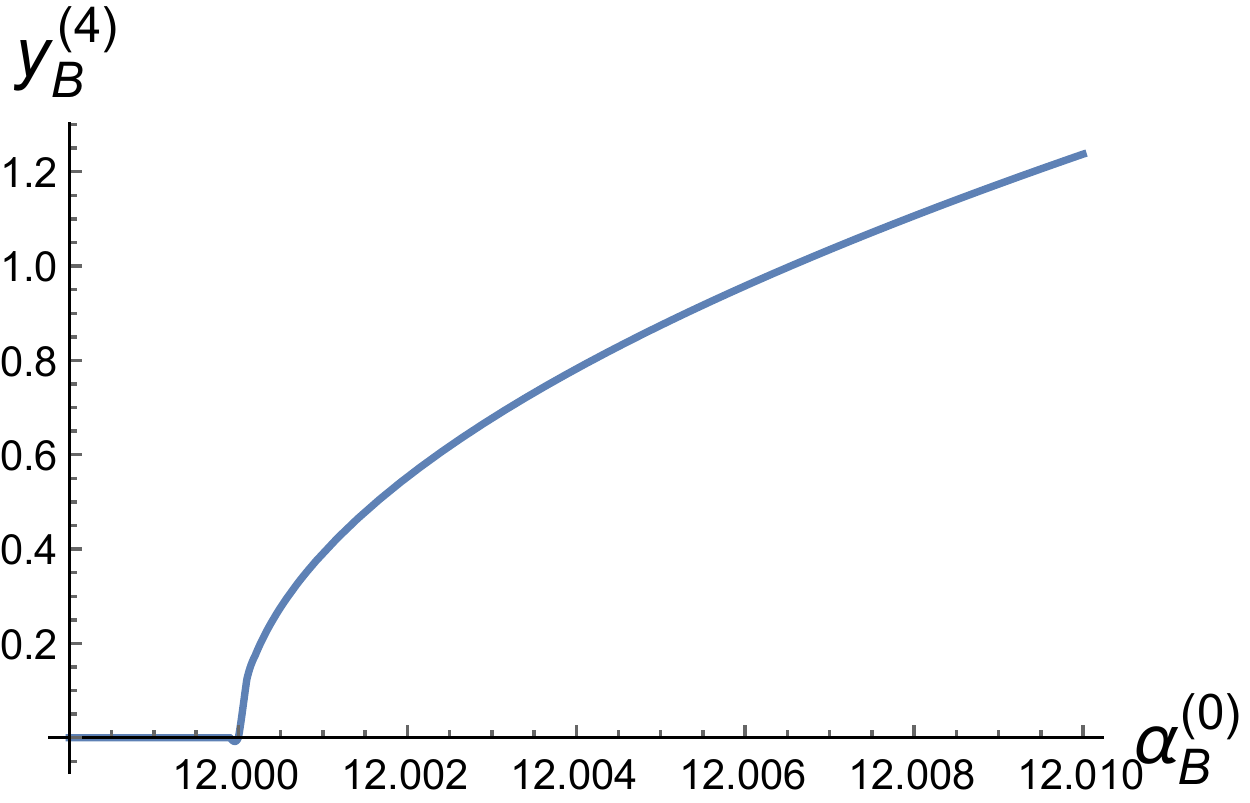}
\caption{
Scalar condensate (left) and charge density (right) as functions of the chemical potential. The last graph is a zoom of the first graph near the critical value of the chemical potential. The relations between the dimensionless quantities in the figures and the physical ones are given by \eqref{eq:vevs}.
}
\label{figcharge}
\end{figure}

The coefficient $y_B^{(4)}$ is proportional to the the expectation value of the charged operator $ \vev{Q\bar{A} X_{5} Q}$, while  $\alpha_B^{(2)}$ is proportional to the baryon charge density $\vev{J^0_B}$. The expectation values of dual operators can be computed by following the usual AdS/CFT prescription. The variation of the on-shell action gives a total derivative term
\begin{equation}
\delta S_{D7}=   \lim_{\rho \to \infty}- 2T_7 \int d^4 x \rho^3 \Bigg[ \frac{\lambda^2}{2}  \Big( \partial_\rho\sigma^\dagger \delta\sigma+\partial_\rho\sigma \delta\sigma^\dagger \Big) -2\pi^2\lambda^2  \partial_\rho \cV_0 \delta\cV_0  \Bigg].
\end{equation}
We can identify the leading terms in the expansions of $\delta \sigma$ and $\delta \tilde{\cV}_0$ as variations of sources for the dual fields, $\delta J$ and $\delta \mu$, and the subleading terms will determine the expectation values
\begin{equation}
\delta \sigma =\frac{1}{R} \Bigg[ \frac{\rho}{L}\delta J+\frac{L^3}{\rho^3}R^4\delta\sigma^{(4)} \Bigg]   , \ \ \delta \cV_0=\delta\mu+\frac{L^2}{\rho^2} R^2\delta \cV_0^{(2)} .
\end{equation}
Then, the identification of the on-shell gravity action with the generating functional of the dual field theory leads to the following expressions for the expectation value of the dual operators
\begin{equation}
\vev{Q\bar{A} X_{5} Q}=\frac{\delta S_{D7}}{\delta J} = \lim_{\rho \to \infty}- T_7  \frac{\lambda^2}{R}\frac{\rho^4}{L} \partial_\rho\sigma, \ \ \vev{J^0_B}= \frac{\delta S_{D7}}{\delta \mu}=\lim_{\rho \to \infty} 4\pi^2T_7  \lambda^2  \rho^3  \partial_\rho \cV_0.
\end{equation}
The relation between the coefficients of the numerical calculation and physical observables in the field theory is
\begin{equation}\label{eq:vevs}
\mu=\frac{L}{\lambda}+\frac{\lambda L}{4 R^4}\alpha_B^{(0)},\ \ \vev{Q\bar{A} X_{5} Q} = \frac{3\pi}{2\sqrt{2}}\frac{\lambda^3 L^3}{R^5} T_7  y_B^{(4)},\ \ \vev{J^0_B}=-\frac{2\pi^2\lambda^3 L^3}{R^4} T_7 \alpha_B^{(2)}.
\end{equation}
To this order in $\lambda$, the critical value of the chemical potential \eqref{eq:muc} corresponds to $\alpha_B^{(0)}=12$. We plot the numerical values in Figure~\ref{figcharge}. For values of the chemical potential below the critical value, the condensate and the charge density are zero. At the critical chemical potential there is a second order phase transition of mean field type to the spontaneously broken phase and the condensate and the charge density increase monotonically as the value of the chemical potential increases. We do not observe any special feature at $\alpha_B^{(0)}=24$, which would correspond to the critical value for  $n=1$ in \eqref{eq:mus5}, but it is possible that a second solution, not connected to the one we are computing, exists after this point.

%
%

\subsection{Free energy and thermodynamics}

\begin{figure}[t] 
\centering
\includegraphics[width=8cm]{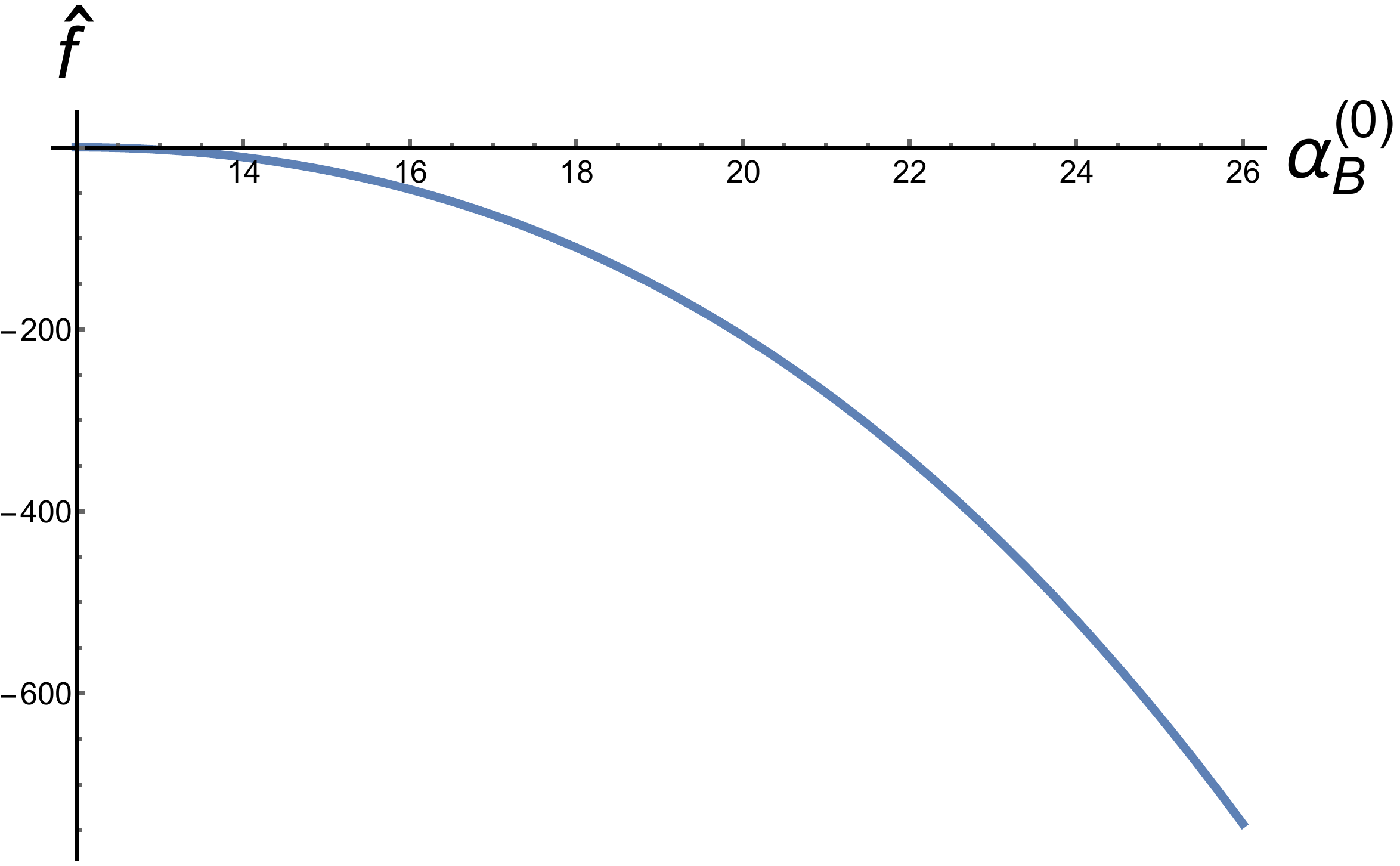}
\caption{Free energy as a function of the chemical potential. The relation between the dimensionless quantities in the figure and physical ones is given by \eqref{eq:vevs},\eqref{eq:f}.
}
\label{figenergy}
\end{figure}

The free energy density is equal to minus the on-shell action density (in Lorentzian signature). Although the on-shell action is divergent, the divergence can be removed by adding a suitable counterterm proportional to the volume of the brane or by subtracting with respect to a reference state, that we can take to be the zero charge density state. In this case, the finite free energy density is\footnote{In general more counterterms are required to regularize the on-shell action, but in the case at hand where all the solutions are normalizable we can neglect this issue.}
\begin{equation}
\cF=2T_7 \int_0^\infty d\rho \Bigg[ \cL_\cV+ \sum_{\underset{odd}{\small \ell \geq 1}} \sum_{a=1}^{N_S^\ell}   \cL_{D7}^{\ell\, a} \Bigg].
\end{equation}
Integrating by parts the action for $\sigma$ and using the equations of motion, one finds that the only non-vanishing contribution is proportional to the action of the gauge field
\begin{equation}
\cF=-2\pi^2\lambda^2T_7  \int_0^\infty d\rho \rho^3  \Big( \partial_\rho \cV_0 \Big)^2 .
\end{equation}
Changing variables to the dimensionless coordinates and numerical functions, the free energy density becomes
\begin{equation}\label{eq:f}
\cF=-\frac{\pi^2\lambda^4 L^4}{8R^8} T_7  \int_0^\infty dx x^3  \big( \partial_x \alpha_0 \big)^2\equiv \frac{\pi^2\lambda^4 L^4}{8R^8} T_7 \hat{f} .
\end{equation}
We have plotted the dimensionless value $\hat{f}$ in Figure~\ref{figenergy}. As we can see, the behavior is consistent with having a second order phase transition from the zero charge phase to the spontaneously broken phase as the value of the chemical potential surpasses the critical value. 

%
%

\begin{figure}[t] 
\centering
\includegraphics[width=8cm]{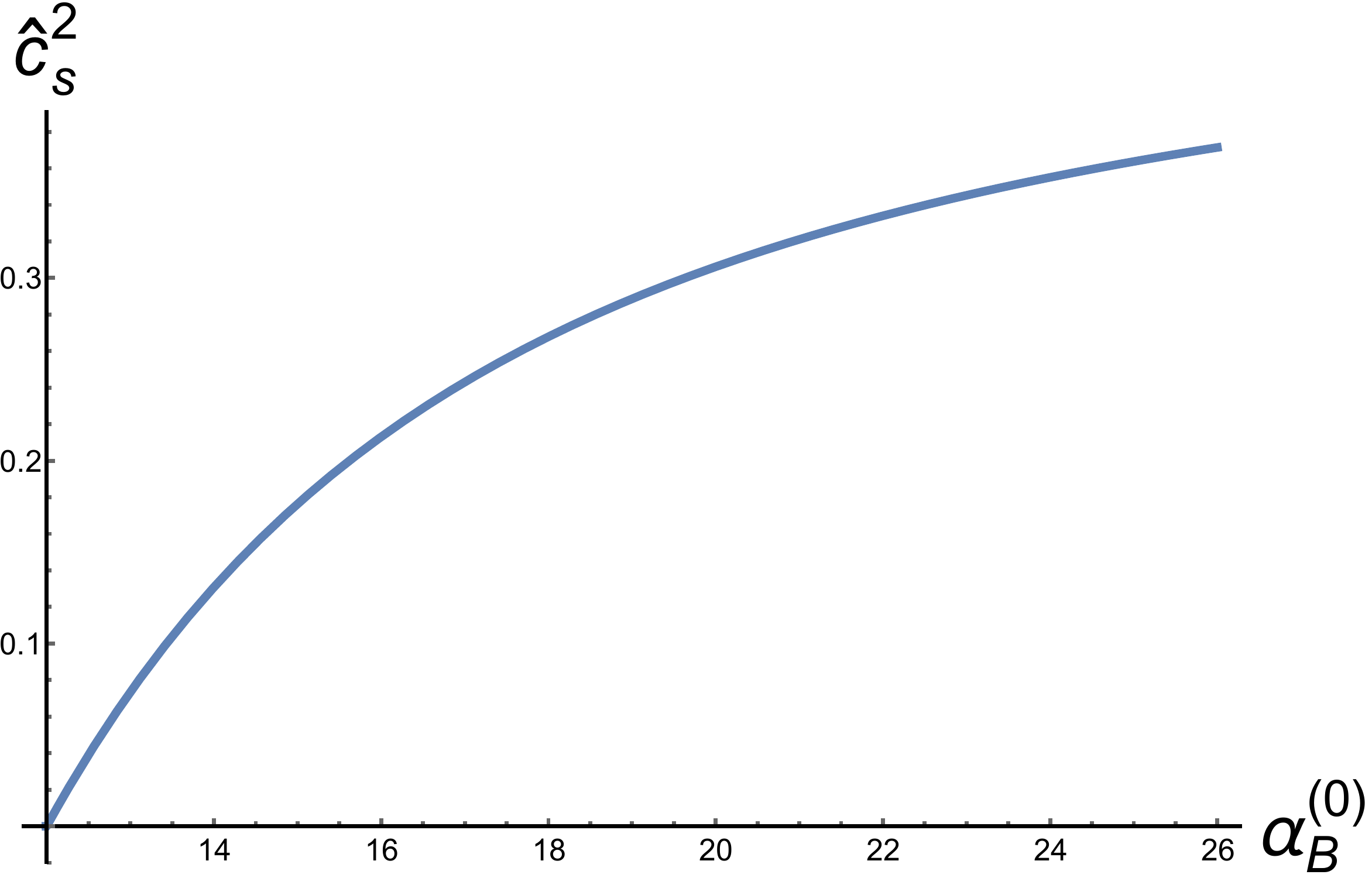}
\caption{Speed of sound as a function of the chemical potential. The relation between the dimensionless quantities in the figure and physical ones is given by \eqref{eq:vevs},\eqref{sspeed}.
}
\label{figspeed}
\end{figure}

The free energy density is equal to minus the pressure $p=-\cF$, and the charge density is the derivative with respect to the chemical potential $n=\frac{\partial p}{\partial\mu}$. The thermodynamic energy density is the Legendre transformation of the free energy density with respect to the chemical potential, thus
\begin{equation}
\varepsilon=\mu n-p.
\end{equation}
From these expressions one finds that the speed of sound is 
\begin{equation}
c_s^2=\frac{\partial p}{\partial \varepsilon}=\frac{n}{\mu \frac{\partial n}{\partial \mu}}.
\end{equation}
The speed of sound measures the stiffness of the equation of state, or in other words, how difficult it is to compress to smaller volumes the matter in nonzero charge phase. We can identify the thermodynamic charge density with $n=\vev{J_B^0}$ in \eqref{eq:vevs}. Then, in terms of the coefficients computed through the numerical solutions:
\begin{equation}\label{sspeed}
c_s^2\simeq \frac{\lambda^2}{4 R^4} \frac{a_B^{(2)}}{\left(\partial a_B^{(2)}/\partial a_B^{(0)}\right)}\equiv \frac{\lambda^2}{4 R^4}\hat{c}^2_s.
\end{equation}
The speed of sound is parametrically small, so the spontaneously broken phase is very soft. At the transition the speed of sound goes to zero. We have plotted the coefficient $\hat{c}_s^2$ in Figure~\ref{figspeed}.

%
%

\section{Summary and outlook}\label{sec:conclusions}

Light baryon operators exist in the large-$N_c$ limit if the $N_c=3$ anti-fundamental representation is identified with the two-index antisymmetric representation. In  \cite{HoyosBadajoz:2009hb}  a string theory construction was proposed as a holographic dual of a concrete large-$N_c$ theory with light baryons. The model consists of the near-horizon geometry of $N_c$ $D3$ branes on a $\mathbb{Z}_2$ orbifold together with an orientifold $O7$ plane and $N_f$ probe (flavor) $D7$ branes. A set of modes of the fields on the $D7$ branes are dual to the light baryon operators. Using the string theory model, we have shown that above a critical baryon chemical potential of the order of the quark mass, there will be a transition from the vacuum to a novel phase with spontaneously broken baryon symmetry, a baryon superfluid. The new phase would be realized by having a non-trivial profile of the modes on the $D7$ branes dual to the light baryon operators. By restricting to a subsector where a single scalar operator condenses and the quark mass is $m_{q}\sim O\left(\lambda_{YM}^0\right)$, we were able to find a solution for the dual fields on the brane and compute the thermodynamic properties of the superfluid phase. An analogous superfluid phase exists for isospin charge in the usual models with flavor branes \cite{Aharony:2007uu}.

There are many possible extensions of this work. An open question is what is the true ground state of the theory above the critical chemical potential or when the distance between the D3 and D7 branes is not of the order of string length. We have studied only the formation of a homogeneous condensate of a single scalar operator, but there are many others that can condense at the same value of the chemical potential, even some vector operators. The state that minimizes the free energy could be a combination of $s$- and $p$-wave components and there could be some degeneracy. It could also happen that the homogeneous state becomes unstable and some spatial symmetries break in the ground state, as has been found in some instances in the Sakai-Sugimoto model \cite{Ooguri:2010xs,Bayona:2011ab,BallonBayona:2012wx}. As the value of the chemical potential is increased, more operators would be susceptible to form a condensate. 

A natural next step is to consider nonzero temperature. Even for small temperatures compared to the quark mass, the phase diagram will be affected because the superfluid baryon phase described here will compete with a phase where quarks do not form bound states, corresponding to  $D7$ branes falling into the horizon. The baryon charge can be at the horizon, in which case it is associated with deconfined quarks in the field theory dual, or it could also be carried by the modes dual to the light baryons, giving a superfluid component. There can also be a competition with phases with an isospin superfluid component due to the formation of a meson condensate, as the ones found in \cite{Erdmenger:2008yj,Ammon:2008fc,Ammon:2009fe}.

Further physical quantities that can be of interest are correlation functions in the charged phase and in particular transport coefficients and the dispersion relation of collective modes. It may also be possible to derive an effective action for the superfluid or a hydrodynamic description of the mixed phase with no bound states of quarks.

A short and non-exhaustive list of interesting generalizations of this work would be: 
\begin{itemize}
\item To go beyond the limit of small number of flavors, i.e. to take $N_f/N_c\sim O(1)$. This requires considering the backreaction of the flavor branes on the geometry, which is a difficult task, but that can be made tractable by ``smearing'', so the individual branes are replaced by a continuous distribution \cite{Casero:2006pt}. This approach has already been succesfully applied to finite density configurations in \cite{Bigazzi:2011db,Bigazzi:2011it,Cotrone:2012um,Bigazzi:2013jqa,Bigazzi:2014qsa}.
\item To look for other stringy constructions with light baryon operators, in particular in geometries which are dual to theories with confinement, such as Klebanov-Strassler \cite{Klebanov:2000hb}. 
\item On a different note it could be possible to construct topological superfluids along the lines of the holographic topological insulators studied in \cite{HoyosBadajoz:2010ac,Ammon:2012dd}.
\end{itemize}
We hope to be able to develop these possible directions and others in the future.\\

\bigskip
\bigskip
\leftline{\bf Acknowledgements}
\smallskip
We would like to thank Johanna Erdmenger, Niko Jokela, Andreas Karch and Aleksi Vuorinen for reading the manuscript and providing useful comments.
G.I. is supported by the FAPESP grants 2016/08972-0 and 2014/18634-9.  C.H. and O.V.  are partially supported by the Ramon y Cajal fellowship RYC-2011-07593, the Asturian grant FC-15-GRUPIN14-108 and the Spanish national grant MINECO-16-FPA2015-63667-P. G.I. and O.V. are also supported by the European CIG grant UE-14-GT5LD2013-618459.

\appendix

\section{Explicit form of the orbifold projection}\label{app:orbifold}

A $S^3$ is a hypersurface in $\mathbb{R}^4$ determined by the equation 
\begin{equation}
x_1^2+x_2^3+x_3^2+x_4^2=1.
\end{equation}
The toroidal/Hopf parametrization is
\begin{equation}
\begin{split}
x_1= &\cos\chi \cos\theta,
\\
x_2= &\cos\chi\sin\theta,
\\
x_3=&\sin\chi\cos\varphi,
\\
x_4=&\sin\chi\sin\varphi.
\end{split}
\end{equation}
With $\chi\in [0,\pi/2]$, $\theta,\varphi\in [0,2\pi]$. We can group the $x_i$ in complex coordinates
\begin{equation}
w=x_1+ix_2=\cos\chi e^{i\theta}, \ \ z=x_3+ix_4=\sin\chi e^{i\varphi},
\end{equation}
so that the equation for $S^3$ is equivalent to the condition $\det U=1$ for the $SU(2)$ matrix
\begin{equation}
U=\left( 
\begin{array}{cc}
w & i z \\
i\bar{z} & \bar{w}
\end{array}
\right).
\end{equation}
The $\det U=1$ condition is invariant under $SU(2)_L\times SU(2)_R$ transformations
\begin{equation}
U\longrightarrow g_L U g_R.
\end{equation}
Note that the $U(1)_L$ rotation $g_L=e^{i\alpha_L \sigma^3}$ shifts
\begin{equation}
\theta\to \theta+\alpha_L, \ \ \varphi\to \varphi+\alpha_L,
\end{equation}
while the $U(1)_R$ rotation  $g_R=e^{i\alpha_R \sigma^3}$ shifts
\begin{equation}
\theta\to \theta+\alpha_R, \ \ \varphi\to \varphi-\alpha_R.
\end{equation}
The $\mathbb{Z}_2\subset SU(2)_L$ subgroup that we use to do the orbifold consists of the identity and the $U(1)_L$ transformation with $\alpha_L=\pi$.

Before the orbifold, the fields on the brane are periodic in the $(\theta,\varphi)$ directions if we are not considering non-trivial fiber bundles. However, these are not the right boundary conditions for the fields that survive the projections. Let us do a large gauge transformation (periodic up to an element of the center of $SU(2N_f)$) on the torus $\gamma(\theta,\varphi)$
\begin{equation}
\begin{split}
\gamma(\theta+2\pi,\varphi)=&\gamma(\theta,\varphi) e^{2\pi i \frac{n_\theta}{2 N_f}},\\[5pt]
\gamma(\theta,\varphi+2\pi)=&\gamma(\theta,\varphi) e^{2\pi i \frac{n_\varphi}{2 N_f}}.
\end{split}
\end{equation}
Where $n_\theta, n_\varphi =0,1,\dots,2N_f-1$.  The fields will be
\begin{equation}
\begin{split}
A_M \longrightarrow &A_M^\gamma =\gamma A_M\gamma^{-1},\\[5pt]
\Phi \longrightarrow & {\Phi^\gamma} =\gamma \Phi \gamma^{-1},\\[5pt]
A_A \longrightarrow &A_A^\gamma =\gamma A_A\gamma^{-1}-i\gamma\partial_A\gamma^{-1}.
\end{split}
\end{equation}
We can expand the $A_M$, $\Phi$ and $A_A$ that are sandwiched between the $\gamma$s in spherical harmonics as usual.  The orbifold projection can be formulated as the periodicity condition (for $X=A_M,A_A,\Phi$)
\begin{equation}
X^\gamma(\theta+\pi,\varphi+\pi)=X^\gamma(\theta,\varphi).
\end{equation}
The form of the projection \eqref{eq:projections} imposes the following condition:
\begin{equation}
\gamma(\theta+\pi,\varphi+\pi)=\gamma(\theta,\varphi) \gamma_7.
\end{equation}
A set of possible transformations labelled by $n=0,1,\dots,N_f$  are
\begin{equation}
\gamma_{n,\pm}(\theta,\varphi)=\left( 
\begin{array}{cc}
 e^{\frac{i}{2 N_f}(n\theta+(N_f-n)\varphi)}\mathbb{1}_{N_f} & \\
  & e^{\frac{i}{2N_f} (n\theta+(N_f-n)\varphi) -\frac{i}{2}(\theta+\varphi \pm(\theta-\varphi)}\mathbb{1}_{N_f}
\end{array}
\right).
\end{equation}
Note that
\begin{equation}
-i\gamma_{n,\pm}^{-1}\partial_\theta\gamma_{n,\pm} =
\left( 
\begin{array}{cc}
 \frac{n}{2 N_f}\mathbb{1}_{N_f} & \\
  & \left( \frac{n}{2N_f}-\frac{1 \pm 1}{2}\right)\mathbb{1}_{N_f}
\end{array}
\right), \ \  
-i\gamma_{n,\pm}^{-1}\partial_\varphi\gamma_{n,\pm} =
\left( 
\begin{array}{cc}
 \frac{N_f-n}{2 N_f}\mathbb{1}_{N_f} & \\
  & \left( \frac{N_f -n}{2N_f}-\frac{1 \mp 1}{2}\right)\mathbb{1}_{N_f}
\end{array}
\right), \ \  
\end{equation}
If we impose that these matrices have the structure of the projected fields compatible with the orientifold projection in \eqref{eq:projectedfields}, we are restricted to two possibilities
\begin{equation}
\gamma_{N_f,+} =\left(\begin{array}{cc}
 e^{\frac{i\theta}{2}}\mathbb{1}_{N_f} & \\
  & e^-{\frac{i\theta}{2}}\mathbb{1}_{N_f}
\end{array}
\right), \ \
\gamma_{0,-} =\left(\begin{array}{cc}
 e^{\frac{i\varphi}{2}}\mathbb{1}_{N_f} & \\
  & e^-{\frac{i\varphi}{2}}\mathbb{1}_{N_f}
\end{array}
\right).
\end{equation}

\section{Projected form of covariant derivatives and commutators}\label{app:projaction}

Let us compute the covariant derivative of block diagonal fields $X_d$. In the following we will denote $2 N_f\times 2N_f$ fields with hats, $\hat{X}_d$ and the $N_f\times N_f$ without hats, $X_d$. Then,
\begin{equation}
\hat{D}_M \hat{X}_d= \partial_M \hat{X}_d+i[\hat{A}_M,\hat{X}_d] =
\left(
\begin{array}{cc}
D_M X_d & \\
 & -(D_M X_d)^*
\end{array}
\right)
\end{equation}
Where
\begin{equation}
D_MX_d=\partial_M X_d+i[A_M,X_d].
\end{equation}
The covariant derivative of block off-diagonal fields $X_o$ is
\begin{equation}
\hat{D}_M \hat{X}_o= \partial_M \hat{X}_o+i[\hat{A}_M,\hat{X}_o] =
\left(
\begin{array}{cc}
 &  D_M X_o \\
-(D_M X_o)^* & 
\end{array}
\right)
\end{equation}
Where
\begin{equation}
D_MX_o=\partial_M X_o+i\left(A_M X_o+X_o A_M^* \right).
\end{equation}
Note that $X_o^T=-X_o$ implies $(D_MX_o)^T=-D_M X_o$.

We will also need the commutator between two fields. For the block diagonal fields
\begin{equation}
i[\hat{X}_d^1,\hat{X}_d^2] =
\left(
\begin{array}{cc}
i[X_d^1,X_d^2] &   \\
 & i \Big( [ X_d^1,X_d^2] \Big)^*
\end{array}
\right).
\end{equation}
This preserves the same block diagonal structure. For the block off-diagonal fields
\begin{equation}
i[\hat{X}_o^1,\hat{X}_o^2] =
\left(
\begin{array}{cc}
 i \Big[X_o^2 (X_o^1)^*-X_o^1 (X_o^2)^* \Big] &   \\
&  i \Big[ (X_o^2)^*X_o^1- (X_o^1)^*X_o^2 \Big] 
\end{array}
\right).
\end{equation}
This also has the same block diagonal structure. Using $X_o^T=-X_o$, we can write it as
\begin{equation}
i[\hat{X}_o^1,\hat{X}_o^2] =
\left(
\begin{array}{cc}
 i \Big[ X_o^1 (X_o^2)^\dagger-X_o^2 (X_o^1)^\dagger \Big] &   \\
&  i \Big[  (X_o^1)^\dagger X_o^2 -(X_o^2)^\dagger X_o^1 \Big] 
\end{array}
\right).
\end{equation}

Since the structure of $2N_f\times 2N_f$ matrices is preserved by the covariant derivative and the commutator, all the terms in the action can be written as the product of two block diagonal or two block off-diagonal matrices. The product of two block diagonal matrices is
\begin{equation}
\hat{X}_d^1\hat{X}_d^2 =
\left(
\begin{array}{cc}
X_d^1 X_d^2 & \\
 & (X_d^1)^T (X_d^2)^T
\end{array}
\right)\, ,
\end{equation}
where we have used that $X_d^\dagger=X_d$. 

The product of two block off-diagonal is
\begin{equation}
\hat{X}_o^1\hat{X}_o^2 =
\left(
\begin{array}{cc}
X_o^1 (X_o^2)^\dagger & \\
 & (X_o^1)^\dagger X_o^2
\end{array}
\right)\, ,
\end{equation}
where we have used that $X_o^T=-X_o$.

In each case the trace becomes
\begin{equation}
\begin{split}
&{\rm Tr}_{2N_f} \Big( \hat{X}_d^1\hat{X}_d^2 \Big)=\,{\rm Tr}_{N_f} \Big[ X_d^1 X_d^2+(X_d^1)^T (X_d^2)^T  \Big]=2\,{\rm Tr}_{N_f} \Big( X_d^1 X_d^2 \Big),\\[5pt]
&{\rm Tr}_{2N_f} \Big( \hat{X}_o^1\hat{X}_o^2 \Big)=\,{\rm Tr}_{N_f} \Big[ X_o^1 (X_o^2)^\dagger+(X_o^1)^\dagger X_o^2  \Big]=\,{\rm Tr}_{N_f} \Big[ X_o^1 (X_o^2)^\dagger +h.c. \Big],\\
\end{split}
\end{equation}
where in both cases we have used the cyclic property of the trace and for the block diagonal fields ${\rm Tr}\, X_d^T={\rm Tr}\, X_d$.


\bibliographystyle{JHEP}

\bibliography{biblio}

\end{document}